\documentclass[%
superscriptaddress,
nofootinbib,
 amsmath,amssymb,
 aps,
prd,
]{revtex4-2}

\usepackage{graphicx}% Include figure files
\usepackage{dcolumn}% Align table columns on decimal point
\usepackage{bm}% bold math
\usepackage{hyperref}% add hypertext capabilities
\usepackage{footmisc}
\usepackage{xcolor}
\newcommand{\qu}[1]{``#1''} % quotation marks

\usepackage[capitalize]{cleveref}
\crefname{equation}{Eq.}{Eqs.}
\crefname{figure}{figure}{figures}
\crefname{table}{table}{tables}
\crefname{subequation}{Eqs.}{Eqs.}
\labelcrefformat{subequation}{#2(#1)#3}
\crefformat{subequations}{#2(#1)#3}
\crefname{section}{section}{sections}
\crefname{appendix}{appendix}{appendices}

\usepackage{siunitx}
\let\svqty\qty
\usepackage{physics}
\let\qty\svqty
\usepackage{amsmath,amssymb,amsfonts}
\def\be#1\ee{\begin{align}#1\end{align}} 
\def\bse#1\ese{\begin{subequations}#1\end{subequations}}

\usepackage{graphicx}
\usepackage{caption}
\usepackage{subcaption}
\usepackage{mathrsfs}

\usepackage{comment}
\usepackage{soul}

\begin{document}

\preprint{APS/123-QED}

%\title{Particle creation with general dispersion relations}% Force line breaks with \\
%\title{Hawking temperature in analogue gravity:\\ reconciling the Bogoliubov and tunneling approaches}
\title{Hawking radiation with dispersion:\\ reconciling the Bogoliubov and tunneling approaches}
\author{Francesco Del Porro}
\email{francesco.del.porro@nbi.ku.dk}
\affiliation{Center of Gravity, Niels Bohr Institute, Blegdamsvej 17, 2100 Copenhagen, Denmark}
\affiliation{Niels Bohr International Academy, Niels Bohr Institute, Blegdamsvej 17, 2100 Copenhagen, Denmark}

\author{Stefano Liberati}%
\email{liberati@sissa.it}
\affiliation{SISSA, International School for Advanced Studies, via Bonomea 265, 34136 Trieste, Italy}
\affiliation{INFN, Sezione di Trieste, via Valerio 2, 34127 Trieste, Italy}
\affiliation{IFPU, Institute for Fundamental Physics of the Universe, via Beirut 2, 34014 Trieste, Italy}

\author{Marc Schneider}
\email{marc.schneider@ehu.eus}
\affiliation{Department of Physics and EHU Quantum Center, University of the Basque Country UPV/EHU,
Barrio Sarriena s/n, Leioa 48940, Spain}

\date{\today}% It is always \today, today,

\begin{abstract}

We investigate Hawking-like particle production in analogue gravity systems with superluminal modified dispersion relations. For a broad class of even, convex, and polynomially bounded dispersion relations, we show that the relevant outgoing modes are governed by an effective horizon induced by dispersive propagation. Extending the near-horizon S-matrix method beyond the purely sonic regime, we compute the Bogoliubov coefficients and demonstrate that, in the low-energy and adiabatic limits, they agree with the tunneling result obtained from the approximant ray. In both cases, the emission spectrum is controlled by an effective surface gravity associated to the effective horizon, leading to controlled deviations from exact thermality. Our results establish an analytical connection between the Bogoliubov and tunneling descriptions in dispersive settings and clarify the conditions under which Hawking radiation remains robust against ultraviolet modifications, with implications extending beyond analogue gravity.
\end{abstract}

\maketitle

%\tableofcontents

\section{Introduction}
The discovery of Hawking radiation from black holes marked a pivotal milestone in our understanding of gravity and its intricate relationship with quantum physics~\cite{hawking1975particle}. This phenomenon has not only deepened our insight into black hole thermodynamics, but has also raised profound questions, such as the information loss problem and the trans-Planckian issue. 

The latter concerns the apparent dependence of Hawking radiation on the ultraviolet (UV) completion of quantum field theory and on the underlying structure of spacetime. Missing a full theory of quantum gravity it seemed for many years that this question would have had to remain unsolved. However, the dawn of analogue gravity, dating back to 1981~\cite{Unruh:1981cg}, marked an unexpected development by providing a framework able to simulate phenomena of quantum field theory in curved spacetime within laboratory settings. Moreover, it provided a concrete example in which the UV completion of the theory was explicitly known.

Indeed, about ten years after~\cite{Unruh:1981cg}, it was realised in~\cite{Jacobson:1991gr} that analogue gravity could provide a physical model for the ``trans-Planckian modes'' believed to be relevant for the Hawking effect. Shortly thereafter, the study of Hawking radiation in the presence of modified dispersion relations was explored further~\cite{Jacobson:1993hn, Unruh:1995je}.

It was soon recognised that analogue gravity systems (see~\cite{Barcelo:2005fc, Schutzhold:2025qna} for extensive reviews) provide an ideal testing ground for the robustness of Hawking radiation against high-energy modifications. This is due to their theoretical simplicity and versatility, as well as their ability to offer explicit tabletop experimental settings in which such predictions can be tested.

Following the aforementioned pioneering works, the robustness of Hawking radiation in analogue systems was later investigated and confirmed by numerous theoretical studies (see, e.g.,~\cite{Unruh:1995je, Brout:1995wp, Corley:1996ar, Corley:1997pr, Himemoto:1999kd, Saida:1999ap, Unruh:2004zk}). 

Nonetheless, it was mainly through the exhaustive investigations carried out by Parentani and collaborators in the second decade of this century that a more comprehensive understanding of this phenomenon was achieved~\cite{Macher:2009tw, Macher:2009nz, Finazzi:2010yq, Finazzi:2011jd, Coutant:2011in, Finazzi:2012iu, PhysRevD.90.044033,Michel:2015aga}. These theoretical investigations relied predominantly on the Bogoliubov coefficient method, used by Hawking in his original derivation \cite{hawking1975particle}, complemented by semi-analytical and numerical analyses which, while showing the robustness of analogue Hawking radiation, did not provide full analytical understanding of its deviations from exact thermality due to the modified dispersion relation of the treated fields.

In order to allow for a purely analytical treatment, and a better understanding at least of the corrections to the Hawking temperature induced by modified dispersion relations, it was proposed in~\cite{DelPorro:2024tuw} to adopt a tunneling picture of Hawking radiation~\cite{Parikh:1999mf, srinivasan1999particle,Vanzo:2011wq,Senovilla_2014,Giavoni:2020gui}. 

This method employs a geometric-optics approximation to describe the particle-creation process as a quantum tunneling event involving complex paths across the causal boundary. In contrast to the Bogoliubov coefficient approach, no knowledge of the asymptotic boundary conditions of the system is required, which allows particle creation to be studied in an almost entirely local manner, wherever the event may occur in spacetime \cite{moretti2012state}. This property is especially useful for localising effects involving effective horizons, such as those associated with the modified dispersion relations that characterise fields in analogue spacetimes. 

Still, the tunneling method has its own limitations, such as the local nature of the insights it provides, which remain insensitive to propagation effects (e.g.~greybody factors), and the intrinsic assumption of the validity of the Wentzel–Kramers–Brillouin (WKB) approximation, a feature it shares with Bogoliubov-based calculations and which is well established at the horizon even in the presence of modified dispersion relations \cite{Schutzhold:2013mba,DelPorro:2023lbv}.

In this work, we return to this issue and provide a generalisation of the calculation performed in~\cite{Coutant:2011in}. In doing so, we bridge the gap between the Bogoliubov and tunnelling approaches and show that they yield the same corrections to the Hawking temperature. The core of our extension of~\cite{Coutant:2011in} consists in explicitly taking into account that the modes associated with modified dispersion relations do not peel off from the Killing horizon of the metric, but rather from an effective horizon slightly  shifted from the former (at least for low energy modes). This observation was pivotal in the derivation presented in~\cite{DelPorro:2024tuw}, and here we also make explicit the regime under which this effective horizon coincides with that determined by the ``approximate'' trajectory introduced in~\cite{DelPorro:2024tuw}. 

In the end, our investigation not only shows the equivalence, for what concerns the Hawking temperature, of the Bogoliubov and tunneling methods but also highlights the truly crucial requirements for the robustness of Hawking radiation. Most remarkably, it also shows that once the low energy infalling mode is determined this is sufficient to reconstruct, via the approximant method, the trajectories of the Hawking partners. Noticeably, the infalling mode also fixes the global vacuum by just requiring regularity at horizon crossing. Finally, the map between the infalling modes and the Hawking partners does not involve trans-Planckian frequencies, hence one may wonder whether it instead reflects the ingoing-to-outgoing mode conversion mechanism that has sometimes been conjectured to underlie the true origin of Hawking quanta \cite{Jacobson:1996zs}.

The article is organised as follows. In Section~\ref{sec:setup}, we introduce the system of interest. In Section~\ref{sec:tunneling} we revise the tunneling calculation the way it was performed in \cite{DelPorro:2024tuw}, refining its mathematical derivation. In Section~\ref{sec:s-matrix}, we carry out the actual calculation of particle production in our analogue black hole using an improved Bogoliubov method. Finally, in Section~\ref{sec:discussion}, we summarise our results and provide a physical interpretation of them, together with the more general lessons they offer for the robustness of Hawking radiation.

\section{Acoustic black hole geometry and modified dispersion relations}\label{sec:setup}

We begin by introducing the basic ingredients of the setup. Following the notation of~\cite{DelPorro:2024tuw}, we consider a $(1+1)$-dimensional acoustic spacetime described by the metric
\begin{equation}\label{eq:akustischeMetrik}
    \dd s^2=-\dd t^2+(\dd x-v(x)\dd t)^2 \,,
\end{equation}
where $v(x)\leq 0$ is the background flow velocity.\footnote{The sign is chosen accordingly with~\cite{Barcelo_2004_Causal} so to have a direct correspondence with outgoing Painlev\'e--Gullstrand coordinates} Throughout the paper we set the speed of sound to $c_s=1$, as in~\cite{Coutant:2011in,DelPorro:2024tuw}, in order to facilitate the comparison with those works.

The metric~\eqref{eq:akustischeMetrik} is stationary and admits the Killing vector field
\begin{equation}
    \chi^a\partial_a=\partial_t \,.
\end{equation}
The corresponding acoustic Killing horizon is located where $\chi^a$ becomes null, namely
\begin{equation}
   | \chi|^2=0
    \qquad \Longleftrightarrow \qquad
    v(x)^2=1 \,.
\end{equation}
As in~\cite{Coutant:2011in,DelPorro:2024tuw}, we assume that $v(x)$ is monotonically increasing, $ v'(x)\geq 0$, so that the geometry possesses a unique horizon at the point where $v=-1$. For convenience, we place it at $x=0$.
Around such horizon the flow velocity can be expanded as
\begin{equation}
v(x) =-1 + \kappa_\textsc{kh}x+ O(x^2)    
\end{equation}
where we have implicitly defined the surface gravity of the Killing horizon of our metric as $\kappa_\textsc{kh}:=v'(0)$.

On this background we consider a massless test scalar field $\phi$ obeying the modified Klein--Gordon equation
\begin{equation}\label{eq:modified_KG} 
    -(\partial_t+\partial_x v)(\partial_t+v\partial_x)\phi+F^2(\partial_x^2)\phi=0 \,.
\end{equation}
The operator $F^2(\partial_x^2)$ encodes the deviation from relativistic propagation and determines the modified dispersion relation.

Since the system is naturally associated with a moving fluid, it is convenient to introduce the preferred frame defined by the fluid four-velocity $u^a$ and by the spacelike unit vector $s^a$ orthogonal to it:
\begin{equation}\label{eq:vect}
    u^a\partial_a := \partial_t+v\partial_x \,,
    \qquad
    s^a\partial_a := \partial_x \,.
\end{equation}
In this basis, $u^a$ is tangent to the fluid flow, while $s^a$ points along the spatial direction singled out by the medium.

A particle interpretation of the field $\phi$ is obtained in the WKB approximation, where one writes
\begin{equation}\label{eq:WKB}
    \phi=\phi_0\,e^{iI} \,,
\end{equation}
with slowly varying amplitude $\phi_0$ and rapidly varying phase $I$. The one-form
\begin{equation}
    k_a=\partial_a I
\end{equation}
is then interpreted as the wave four-momentum, or equivalently as the gradient of the \qu{point-particle action} $I$.

At leading WKB order, substituting~\eqref{eq:WKB} into~\eqref{eq:modified_KG} yields the local dispersion relation
\begin{equation}\label{eq:local_dispersion}
    \omega^2 = F(k)^2 \,,
\end{equation}
where
\begin{equation}
    \omega:= -k_a u^a \,,
    \qquad
    k:= k_a s^a \,,
\end{equation}
are respectively the energy and spatial momentum measured in the preferred frame. The modified dispersion relation is intended to describe the low-energy sector of the theory, namely momenta below some ultraviolet scale $\Lambda$, and should therefore be understood as an expansion in $k/\Lambda$.

In~\cite{DelPorro:2024tuw} we specialised to the quartic dispersion relation
\begin{equation}\label{eq:MDR}
    F(k)^2 = k^2+\xi\frac{k^4}{\Lambda^2} \,,
\end{equation}
where $\xi=\pm1$ distinguishes between superluminal ($\xi=+1$) and subluminal ($\xi=-1$) propagation.\footnote{The superluminal case corresponds to the famous Bogoliubov dispersion relation for quasi-particles in Bose--Einstein condensates. These are presently the only systems in which the quantum analogue of Hawking radiation has been convincingly observed~\cite{Barcelo:2005fc, MunozdeNova:2018fxv}.}

In the present work we will mostly focus on the superluminal case, while commenting on the subluminal case in the final part of the paper. Unless otherwise stated, however, we shall keep $F$ generic and assume that it satisfies the following properties:
\begin{enumerate}
    \item $F(k)$ is \emph{even}, so that the dispersion relation is invariant under $k\to -k$ in the preferred frame\footnote{More specifically, we have in mind that dispersion relation is CPT-invariant in the preferred frame as in Ho\v{r}ava gravity \cite{Herrero-Valea:2023zex}: a Lorentz violating, power-counting renormalizable, quantum gravity theory, to which the present discussion can be easily exported, see \cite{DelPorro:2023lbv}.}
    \begin{equation}
        F(-k)=F(k) \,.
    \end{equation}
    \item $F(k)$ is \emph{convex}, namely
    \begin{equation}
        F''(k)\geq 0
        \qquad \text{for all } k\in\mathbb{R} \,.
    \end{equation}
    In particular, this excludes additional local minima of the dispersion relation, and hence the presence of extra roton-like branches \cite{Barcelo:2005fc}.
    \item $F(k)$ is \emph{polynomially bounded}, in the sense that there exists some $n\in\mathbb{N}$, with $n\geq 2$, such that
    \begin{equation}
    \biggl|    \lim_{k\to\infty} \frac{F(k)}{k^n} \biggr|<\infty \,.
    \end{equation}
\end{enumerate}

Because the background is stationary, the field equation is invariant under translations generated by $\chi^a$. One can therefore decompose the solutions of~\eqref{eq:modified_KG} into stationary modes of definite Killing energy $\Omega$
\begin{equation}
    \phi_\Omega(x,t)=e^{-i\Omega t}\,\varphi_\Omega(x)\,,
    \qquad
    \varphi_\Omega(x)=e^{iS_\Omega(x)} \,,
\end{equation}
where
\begin{equation}
    \Omega:= -k_a\chi^a
\end{equation}
is the conserved quantity associated with stationarity, and
\begin{equation}
    S_\Omega(x):= I+\Omega t
\end{equation}
is the reduced action~\cite{Coutant:2011in}.

It is useful to make explicit the relation between the conserved Killing energy $\Omega$ and the energy $\omega$ measured in the preferred frame. From~\eqref{eq:vect} one immediately finds
\begin{equation}
    \chi^a=u^a-v(x)s^a \,.
\end{equation}
Contracting this identity with $-k_a$ gives
\begin{equation}
    \Omega=-k_a\chi^a
    =-k_a u^a + v(x)k_a s^a
    =\omega+v(x)k \,.
    \label{eq:Omega}
\end{equation}
Equation~\eqref{eq:Omega} shows that the conserved Killing energy differs from the comoving energy by a Doppler-like shift due to the background flow.

Combining~\eqref{eq:local_dispersion} with~\eqref{eq:Omega}, one obtains the Hamilton--Jacobi form of the field equation,
\begin{equation}\label{eq:MDR_combined}
    \bigl(\Omega-v(x)k\bigr)^2 = F(k)^2 \,.
\end{equation}
For a given value of the conserved Killing energy $\Omega$, this is an algebraic equation for the local momentum $k(x)$. Its different roots describe the distinct WKB branches, i.e.\ the different modes propagating on the acoustic black hole background. We now turn to their analysis.

%%%%%

\subsection{Modes and characteristics} \label{subsec:modes}
At any spatial location $x$, \cref{eq:MDR_combined} can admit more than two solutions, therefore giving rise to a richer structure of modes propagating with respect to the relativistic case $F(k)^2=k^2$.

If $F$ describes a superluminal behaviour, the number of \textit{real} solutions $k(x)$ changes on the two sides of the Killing horizon \cite{DelPorro:2023lbv}. For $x>0$, outside the Killing horizon, one has two solutions, namely an \textit{ingoing} mode $k_->0$ and an \textit{outgoing} one $k_+<0$. These have opposite sign for their group velocity. The ingoing mode has negative group velocity and, in the lab frame $\{t, x \}$, it moves towards the horizon from outside. Conversely, the outgoing solution is associated with a positive group velocity, traveling outwards in the lab frame.

Inside the horizon, the outgoing and the ingoing modes $k_\pm$ are still present. However, in this case, despite having opposite sign in their group velocity, they both will be moving inwards in the lab frame, due to the fact that for $x<0$ a trapped region for modes with energy $\Omega\lesssim \Lambda$ exists.

\begin{figure}[t!]
    \centering
    \includegraphics[scale=0.16]{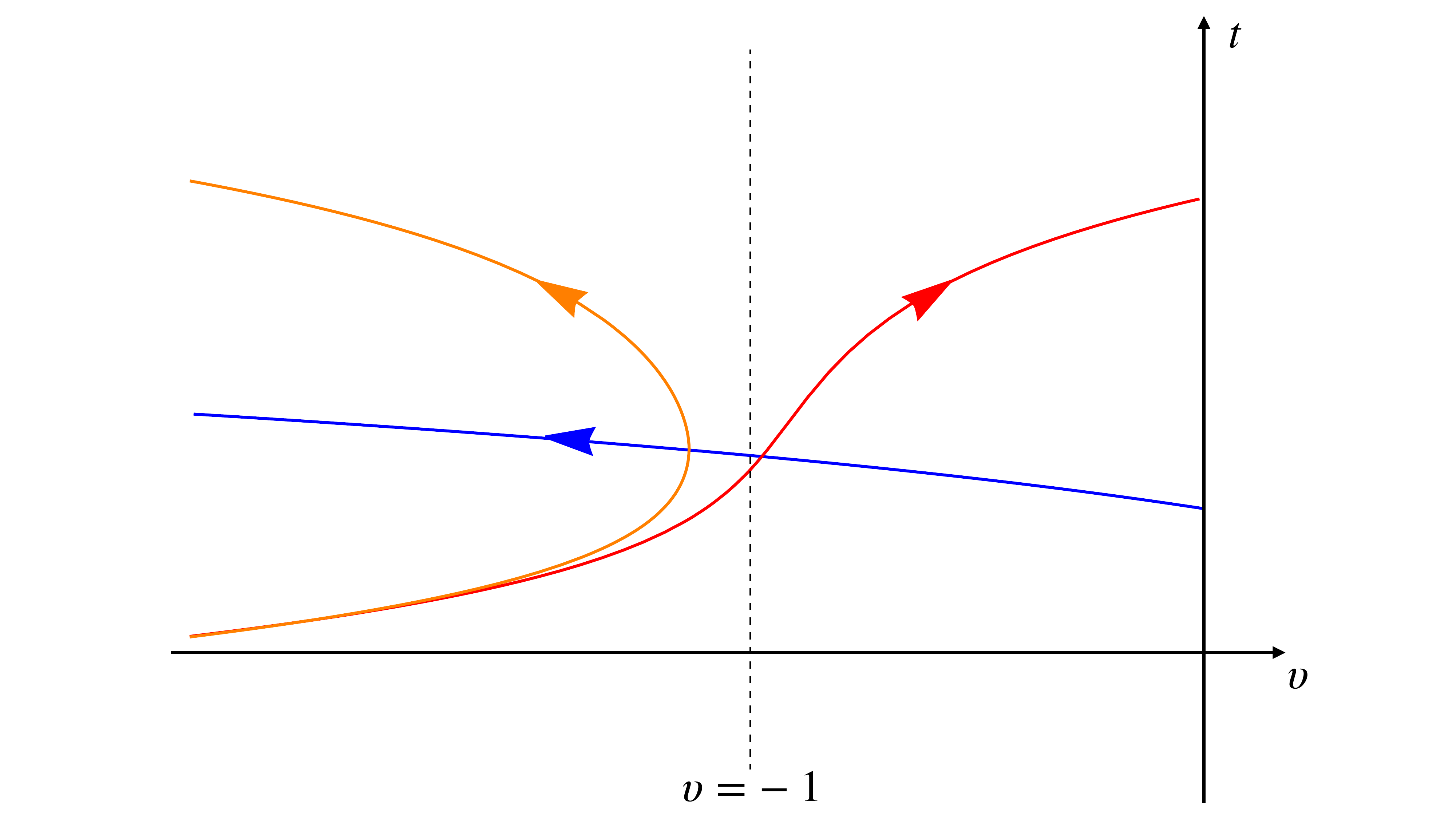}
    \caption{Mode characteristics in the $\{v,t\}$ plane. Starting from the left side of the plot (inside the Killing horizon), at fixed $v$ one can identify four distinct solutions: two soft ones, respectively the blue ($k_-$) and the upper branch of the orange ($k_+$) lines, and two hard ones, namely the red ($k_\Lambda^{\rm red}$) and the lower branch of the orange ($k_\Lambda^{\rm or}$) line, which almost overlap one to another. Moving towards the right in the horizontal axis, $k_+$ and $k_\Lambda^{\rm or}$ degenerate into a single solution, creating a \qu{turning point} for the orange line. The real solutions then reduce to two: the blue line ($k_-$) and the soft red line ($k_+$). All the modes have been taken with Killing energy $\Omega/\Lambda=2 \times10^{-2}$ and $F(k)^2=k^2+k^4/\Lambda^2$.} 
    \label{fig:characteristics}
\end{figure}

Notably, they are accompanied with two other real solutions of order $\Lambda$, which we will name $k_\Lambda^{\rm red}$ and $k_\Lambda^{\rm or}$, for a reason that will become clear later. These solutions of order $\Lambda$ cease to be real approaching the Killing horizon from inside and then become complex conjugated solutions of \cref{eq:MDR_combined}. In particular, this happens when two of the solutions, namely $k_+$ and $k_\Lambda^{\rm or}$, become degenerate. As one can see from Figure \ref{fig:characteristics}, this degeneracy determines a \qu{turning point} $x_{\rm tp}<0$ in the point particle's trajectory, after which we are left with only two real modes. Along the text, we will refer to $k_\Lambda^{\rm red}$ and $k_\Lambda^{\rm or}$ as \textit{hard} solutions, due to their high-energy nature, while $k_\pm$ will be labeled as \textit{soft} solutions.

For each mode, its characteristic is defined through the Hamilton equation
\begin{equation} \label{eq:x_characteristics}
    \frac{{\rm d}t}{{\rm d}x} = \frac{\partial k}{\partial \Omega} = \frac{1}{c_g(k(x,\Omega))+v(x)}\,,
\end{equation}
where the last equality can be computed by differentiating \cref{eq:MDR_combined} by $\Omega$ and
\begin{align} \label{eq:group_velocity}
   c_g(k):= \frac{{\rm d}F (k)}{{\rm d}k} \,
\end{align}
is the group velocity of the ray. The integral lines determined by \cref{eq:x_characteristics} are represented in Figure \ref{fig:characteristics} as curves $(t(v), v)$, to keep the generality about the shape of the profile $v(x)$. This is always possible as long as $v(x)$ has a monotonous behaviour, as we have already assumed in Section \ref{sec:setup}. 

Let us emphasize that \cref{eq:x_characteristics} represents the trajectory followed by a wavepacket in the geometric optic approximation. Indeed, taken a superposition of monochromatic modes $\{\phi_\Omega\}$ peaked on some energy $\Omega_0$ with some distribution $\rho(\Omega,\Omega_0)$, the envelope
\begin{equation}
    \int {\rm d}\Omega \, \rho(\Omega,\Omega_0) e^{-i \int (\Omega {\rm d}t - k_\Omega {\rm d}x)}\,
\end{equation}
can be approximated by its saddle points $\delta_\Omega I=0$, satisfying
\begin{equation}
     \frac{{\rm d}t}{{\rm d}x} = \frac{\partial k}{\partial \Omega} \biggr|_{\Omega_0} \,,
\end{equation}
as in \cref{eq:x_characteristics}. 

\subsection{The approximant}\label{subsec:approximant}
The structure of the modes that emerge from the previous analysis makes clear that, if this black hole is let to evaporate by the Hawking effect, the role of the Hawking pair is played by the two soft outgoing modes with momentum $k_+(x)$ on the two sides of the horizon. Along this paper, we will name the solution with support outside of the Killing horizon as $\varphi^{\rm H}_\Omega(x)$ and its partner (with support inside of the horizon) as $\varphi^{\rm P}_\Omega(x)$. In \cite{DelPorro:2024tuw}, we noticed that the characteristic of these two modes can be mostly described, by the \textit{approximant ray}, instead of \cref{eq:x_characteristics}. This offers an immediate tool to compute the evaporation rate even in case of modified dispersion. In this section, we are going to introduce this concept and revise its mathematical derivation.

Consider $x>0$, the basic idea is to \qu{track} the evolution of $k_+(x)$ in terms that of $k_-(x)$. For a given $\Omega>0$, the two functions $k_\pm(x)$ satisfy the two equations:
\begin{subequations}
    \begin{align}
    &\Omega - v \left( x \right) k_+(x)= F(k_+(x))  \qquad \mbox{outgoing}\,, \label{eq:x_outgoing}\\
    &\Omega - v \left( x \right) k_-(x)= F(k_-(x)) \qquad \mbox{ingoing} \,, \label{eq:x_ingoing}
\end{align}
\end{subequations}
where the sign in front of $F$ is taken accordingly with the fact that, in the limit of a flat background ($v=0$) we need $F=\Omega>0$.

Since $k_-(x)$ is an invertible and continuous function, we can formally solve for $x(k_-)$ and then define the relation 
\begin{equation} \label{eq:x_parametrization}
    k_+(k_-)=:-k_- + R(k_-)=:f(k_-)
\end{equation}
 where the function $R(k_-)$ is taken to go to 0 as $v \to 0$. In what follows, we are going to manipulate \cref{eq:x_outgoing,eq:x_ingoing} to determine the shape of $R(k_-)$. 
 
Taking the difference between Eq.\eqref{eq:x_outgoing} evaluated on $k_+$ and Eq.\eqref{eq:x_ingoing} evaluated on $k_-$ we have
\begin{align} \label{eq:x_outgoing_k-}
   v\left( x \right)(2k_- - R) = F \left(-k_- +R \right) - F \left(k_- \right) \,.
\end{align}
Since $F$ is an even function of $k_-$ by assumption, we can rewrite
\begin{equation} \label{eq:x_r_k-}
  R = \frac{2 v \left(x \right)  k_-}{v \left( x \right) + \frac{F(k_- - R)-F(k_-)}{R}} \,.
\end{equation}
Secondly, the convexity assumption implies that the following inequality always holds
\begin{align} \label{eq:convex_F}
    \frac{F(k_--R)-F(k_-)}{R} \ge-c_g^-(k_-) \,.
\end{align}
Finally, the polynomial boundedness implies that the $j-$th derivative of $F$ ($j \le n$) is bounded by an appropriate power law
\begin{align} \label{eq:poly_bound}
\left| \frac{1}{j!} \frac{{\rm d}^jF(k)}{{\rm d}k^j} \right|:=\frac{|F^{(j)}(k)|}{j!} \le C_j\frac{k^{n-j}}{\Lambda^{n-1}}  \,,
\end{align}
where the $C_j \ge 0$ are dimensionless numbers and the powers of $\Lambda$ have been inserted for dimensional reasons. 
Via Taylor's theorem we have
\begin{align} \label{eq:taylor_F}
F(k_--R)= F(k_-)-c_g(k_-)R + \frac{F^{(2)}(\zeta)}{2}R^2\,,
\end{align}
where the last term represents the Lagrange's form of the remainder with $\zeta \in [k_--R,\, k_-]$. Therefore, putting \cref{eq:convex_F,eq:poly_bound} together we have a double-sided bound
\begin{align} \label{eq:upper_bound}
-c_g(k_-) &\le\frac{F(k_--R)-F(k_-)}{R}\\
&\le  -c_g(k_-)+ \frac{1}{2}C_2 \frac{R^{n-1}}{\Lambda^{n-1}}  \nonumber\,,
\end{align}

In particular we have
\begin{align} \label{eq:both_bounds}
\frac{2 v \left( x \right)  k_-}{v \left( x \right) -c_g(k_-) + \frac{1}{2}C_2 \frac{R^{n-1}}{\Lambda^{n-1}}} \le R  \le \frac{2 v \left( x \right)  k_-}{v \left( x \right) -c_g(k_-) } \,.
\end{align}

In a generic point $x$, the correction proportional to $(R/\Lambda)^{n-1}$ is subdominant with respect to $v(x)-c_g(k_-)$, due to its $\Lambda$-suppression. However, close to
\begin{align} \label{eq:EFH}
    v \left(x \right)-c_g(k_-) =0\,,
\end{align}
the term $(R/\Lambda)^{n-1}$ becomes relevant. We would like to point out that \cref{eq:EFH} has similarly been found in \cite{DelPorro:2024tuw} and, as it will become clear later, its solutions are associated with the presence of an \textit{effective horizon} (EFH) for the approximant. 

Let us now consider a solution of~\cref{eq:EFH} and denote it by $x_0$. Since, in our coordinate system, the Killing horizon is located at $x=0$, the quantity $x_0$ measures the coordinate distance between the Killing horizon and the effective horizon.

Expanding near $x=x_0$, one has at leading order
\begin{equation}
    v(x)-c_g(k_-)
    =
    \mathcal O(x-x_0) \,.
\end{equation}
Hence, close to $x_0$, the lower bound in~\cref{eq:both_bounds} is dominated by the correction term in the denominator, and becomes
\begin{equation}\label{eq:r_bound_x0}
R
    \gtrsim
    -\,\frac{4\bar v\bar k}
    {C_2\,R^{\,n-1}/\Lambda^{\,n-1}}
    +\mathcal O(x-x_0) \,,
\end{equation}
where we introduced
\begin{equation}
    \bar v := v(x_0) \,,
    \qquad
    \bar k := k_-(x_0,\Omega) \,.
\end{equation}
Equation~\eqref{eq:r_bound_x0} can be rearranged into the estimate
\begin{equation}\label{eq:r_scaling_x0}
    R
    \gtrsim
    -\left(\frac{4\bar v\bar k}{C_2}\right)^{1-\gamma}
    \Lambda^\gamma \,,
    \qquad
    \gamma:= \frac{n-1}{n} \,.
\end{equation}

Substituting this estimate back into~\cref{eq:both_bounds}, one obtains
\begin{equation}\label{eq:both_bounds_2}
    \frac{2v(x)\,k_-}
    {v(x)-c_g(k_-)+\left(\dfrac{4C_2\bar v \bar k}{\Lambda}\right)^\gamma}
    \le R \le
    \frac{2v(x)\,k_-}{v(x)-c_g(k_-)} \,.
\end{equation}
Within the assumptions made above, this bound is exact for any superluminal, convex, and polynomially bounded dispersion relation.

The approximant is obtained by working in the regime 
\begin{equation}\label{eq:perturbative_param}
    \left(\frac{4C_2\bar v \bar k}{\Lambda}\right)^\gamma \ll 1 \,.
\end{equation}
This assumption is consistent with our setup. Indeed, $\bar k$ is the momentum of the ingoing branch $k_-(x)$ evaluated at the EFH, and is therefore expected to remain well below the ultraviolet scale $\Lambda$. Unlike the outgoing branch, the ingoing mode does not experience the geometrical blueshift associated with horizon crossing, and thus remains in the low-energy regime along its trajectory. In this sense, solving~\cref{eq:EFH} within the approximation~\eqref{eq:perturbative_param} amounts to neglecting corrections of order $(\bar k/\Lambda)^\gamma$.

Under this assumption, the two bounds in~\cref{eq:both_bounds_2} coincide at leading order and define the approximant relation
\begin{equation}\label{eq:approximant}
    R(x,\Omega)
    =
    \frac{2v(x)\,k_-(x,\Omega)}
    {v(x)-c_g^-(x,\Omega)} \,,
\end{equation}
which, together with~\cref{eq:x_parametrization}, determines the approximant ray. Here we introduced the shorthand
\begin{equation}
    c_g^\pm(x,\Omega):= c_g\!\left(k_\pm(x,\Omega)\right) \,.
\end{equation}
As expected, one has $R\to 0$ in the limit $v\to 0$.

A completely analogous construction applies inside the horizon, namely for $x<0$, where the relevant outgoing branch is the partner mode. In this case, the mode propagating towards the future has negative Killing energy $-\Omega<0$ and negative preferred-frame energy $-F(k)<0$. One must therefore consider the negative-frequency root of~\cref{eq:MDR_combined}. The relevant pair of equations is then
\begin{subequations}\label{eq:inside_branches}
\begin{align}
    -\Omega-v(x)\,k &= -F(k) \,,
    \qquad \text{(outgoing)} \,,
    \label{eq:x_outgoing_inside}
    \\
    \Omega-v(x)\,k &= F(k) \,,
    \qquad \text{(ingoing)} \,.
    \label{eq:x_ingoing_inside}
\end{align}
\end{subequations}
Defining now
\begin{equation}
    k_+(k_-)=k_-+R(k_-) \,,
\end{equation}
and summing~\cref{eq:x_outgoing_inside,eq:x_ingoing_inside}, one finds
\begin{equation}\label{eq:x_outgoing_k-_inside}
    v(x)\,\bigl(2k_-+R\bigr)
    =
    F(k_-+R)-F(k_-) \,.
\end{equation}
This can be rewritten as
\begin{equation}\label{eq:r_k-_2a}
    R
    =
    -\,\frac{2v(x)\,k_-}
    {v(x)-\dfrac{F(k_-+R)-F(k_-)}{R}} \,.
\end{equation}
Since the same assumptions for $F$ still apply, the previous argument can be repeated essentially unchanged. One, therefore, obtains the interior approximant
\begin{equation}\label{eq:approximant_inside}
    R=-\,\frac{2v(x)\,k_-(x,\Omega)}
    {v(x)-c_g^-(x,\Omega)} \,.
\end{equation}
In particular, as $v\to -\infty$ one has $R\to -2k_-$. Equivalently, the outgoing branch asymptotically acquires a group velocity opposite to that of the ingoing one, in agreement with the discussion of~\cite{DelPorro:2024tuw}.

\begin{figure}[t!]
    \centering
    \includegraphics[scale=0.16]{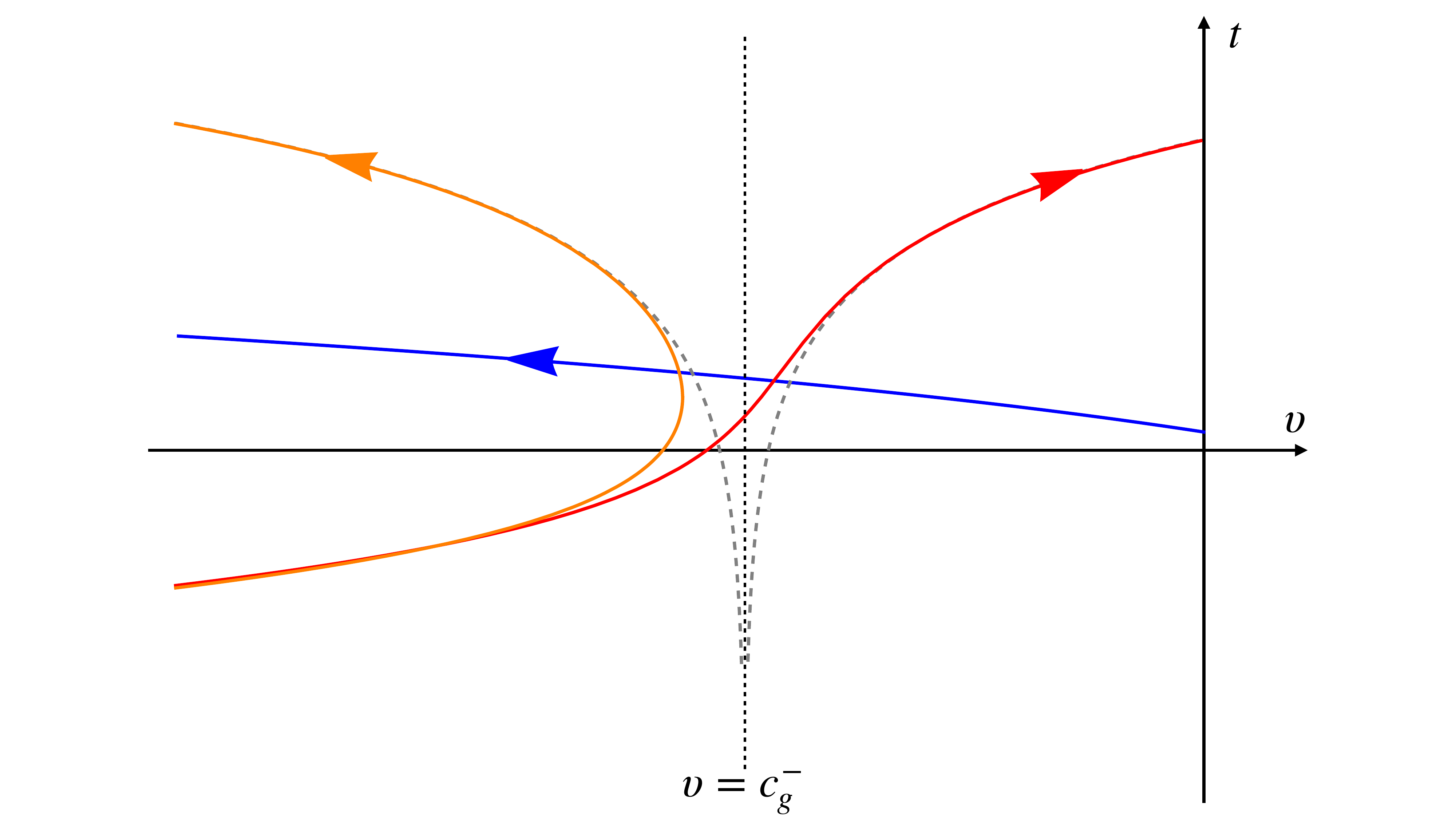}
    \caption{Modes' and approximant's characteristics in the $\{v,t\}$ plane. On top of the characteristics depicted in Figure \ref{fig:characteristics}, the dashed gray lines which peel out from the EFH $\{v=c_g^- \}$ represent the approximant ray. One can notice how the approximant overlaps with the physical trajectories on the two soft branches described by $k_+$ on both sides of the EFH.  All the modes have been taken with Killing energy $\Omega/\Lambda=2 \times10^{-2}$ and $F(k)^2=k^2+k^4/\Lambda^2$.} 
    \label{fig:approximant}
\end{figure}

\subsection{Near-horizon behaviour}

It is particularly instructive to analyse the behaviour of the modes in the neighbourhood of the EFH. Substituting the approximant~\eqref{eq:approximant} into~\eqref{eq:x_parametrization}, one finds
\begin{equation}\label{eq:k_+_approximant}
    k_+(x)=-k_-(x)+\frac{2v(x)\,k_-(x)}{v(x)-c_g^-(x,\Omega)} \,.
\end{equation}
Near the EFH, the second term dominates, since by definition
\begin{equation}
    v(x_0)-c_g^-(x_0,\Omega)=0 \,.
\end{equation}
Expanding the denominator to first order around $x=x_0$, we obtain
\begin{equation}
    v(x)-c_g^-(x,\Omega)\simeq \kappa_\textsc{efh}(\Omega)\,(x-x_0) \,,
\end{equation}
where we introduced the \emph{EFH peeling surface gravity}~\cite{DelPorro:2024tuw},
\begin{equation}\label{eq:Xkappa}
    \kappa_\textsc{efh}(\Omega):=\left.\frac{\dd}{\dd x}\Bigl(v(x)-c_g^-(x,\Omega)\Bigr) \right|_{x=x_0} \,.
\end{equation}
Therefore, to leading order in $x-x_0$,
\begin{equation}\label{eq:x_kappa}
    k_+(x)
    \simeq
    \frac{2\bar v \bar k}{\kappa_\textsc{efh}(\Omega)\,(x-x_0)} \,.
\end{equation}
Since, in the WKB approximation $k_+=\partial_x S_\Omega$, integration yields
\begin{equation}\label{eq:S_log}
    S_\Omega(x)
    \simeq
    \frac{2\bar v\bar k}{\kappa_\textsc{efh}(\Omega)}\,
    \ln (x-x_0) \,.
\end{equation}

Even though in the following we will often suppress the argument of $\kappa_\textsc{efh}(\Omega)$ for notational simplicity, it is important to stress that $\kappa_\textsc{efh}$ is, in general, frequency dependent. This is a direct consequence of the modified dispersion relation: the group velocity $c_g^-(x,\Omega)$ depends on the branch under consideration and therefore on $\Omega$, so the position $x_0$ of the EFH and the corresponding peeling coefficient inherit this dependence. This is consistent with the results already found in~\cite{DelPorro:2024tuw,DelPorro:2023lbv}, and we shall comment further on its implications below.

We proceed with further analysing the relation between the coefficient $2\bar v\bar k$ and the conserved Killing energy $\Omega$. Evaluating the ingoing dispersion relation~\eqref{eq:x_ingoing} at $x=x_0$, one finds
\begin{equation}\label{eq:ingoing_x0}
    \Omega-\bar v\bar k = F(\bar k) \,.
\end{equation}
On the other hand, the defining condition of the EFH reads
\begin{equation}\label{eq:efh_x0}
    \bar v = c_g(\bar k)=F'(\bar k) \,.
\end{equation}
Combining~\eqref{eq:ingoing_x0} and~\eqref{eq:efh_x0}, we obtain the exact identity
\begin{equation}\label{eq:Omega_EFH_correct}
    \Omega
    =
    F(\bar k)+\bar k\,F'(\bar k) \,.
\end{equation}
Therefore, in general,
\begin{equation}\label{eq:Omega_eff_def}
    2\bar v\bar k = 2\bar kF'(\bar k)\,.
\end{equation}
Let us consider first the relativistic limit: in this case $F(k)$ is homogeneous of degree one and hence satisfies $F(k)=kF'(k)$ on each branch so $2\bar v\bar k =\Omega$. However, for a generic modified dispersion relation, this is evidently not true anymore. So the coefficient controlling the logarithmic singularity is not the Killing energy itself, but an effective frequency defined as
\begin{equation}\label{eq:Omega_eff}
    \Omega_{\rm eff}(\Omega):=2\bar v \bar k \,.
\end{equation}

This result might seem bizarre but it can be understood geometrically. In the relativistic case, the wave covector $k_a$ is directly related to the tangent to the ray, so the conserved Killing energy $-k_a\chi^a$ also controls the near-horizon peeling. By contrast, in a dispersive theory the rays are generated by the Hamilton equation, see Eq.~\eqref{eq:x_characteristics}, and their velocity is governed by the group velocity $c_g=F'(k)$ rather than by $k_a$ alone --- see \cite{DelPorro:2025zyv}, Appendix B, and also \cite{Frolov:2012ux,Barcaroli:2015xda} for a more detailed discussion on this point. As a result, the coefficient entering the logarithmic singularity of the WKB phase is not the conserved Killing energy itself, but the combination $\Omega_{\rm eff}=2\bar v \bar k$, which reduces to $\Omega$ only in the relativistic limit.
Note that this subtle point was overlooked in \cite{DelPorro:2024tuw}, however, as we shall see in the end, in the low energy limit $\Omega/\Lambda\ll 1$ it does amount only to a rescaling of the coefficient in front of the $(\Omega/\Lambda)^2$ correction to the temperature found in that work.

Using Eq.~\eqref{eq:Omega_eff}, we rewrite Eq.~\eqref{eq:S_log} as
\begin{equation}
    S_\Omega(x) \simeq \frac{\Omega_{\rm eff}(\Omega)}{\kappa_\textsc{efh}(\Omega)} \ln(x-x_0):=\frac{\Omega}{\kappa_{\rm eff}(\Omega)}\ln(x-x_0) \,,
    \label{eq:S-pole}
\end{equation}
where we have defined 
\begin{equation} \label{eq:kappa_eff}
    \kappa_{\rm eff}(\Omega):= \frac{\Omega}{\Omega_{\rm eff}(\Omega)}\kappa_\textsc{efh}(\Omega)
    \,,
\end{equation}
so to include all the dispersive effects in a single quantity which will characterise our effective temperature.

Therefore, the outgoing exterior mode behaves near the EFH as
\begin{equation}\label{eq:H_mode}
    \varphi_\Omega^{\rm H}(x) \simeq \frac{(x-x_0)^{i\Omega/\kappa_{\rm eff}}}{\sqrt{4\pi\Omega}} \,.
\end{equation}
Likewise, approaching the EFH from the interior, the partner mode is approximated by
\begin{equation}\label{eq:P_mode}
    \varphi_\Omega^{\rm P}(x) \simeq \frac{(x_0-x)^{i\Omega/\kappa_{\rm eff}}}{\sqrt{4\pi\Omega}} \,.
\end{equation}

\section{Particle production via the tunneling method}\label{sec:tunneling}
Now that we have set up the problem and introduced the concept of the approximant ray, let us explain how to use it to compute the black hole's radiative properties. 

The first notable thing is the link of the  concept of the approximant with the presence of an EFH, and a simple-pole structure in the WKB phase $I$. In terms of a particle interpretation (so the trajectories minimizing the action), the approximant describes outgoing rays which \qu{peel-out} exponenitally from the EFH at $x=x_0$.

This resembles exactly the relativistic behaviour, where null, outgoing geodesic would peel-out exponentially from the Killing horizon. 

In fact, the presence of a simple pole in $S_\Omega$ is a smoking gun for Hawking radiation \cite{Parikh:1999mf, Giavoni:2020gui}. Within the tunneling picture, this statement acquires the quantitative form of
\begin{equation}
    \Gamma=e^{-2 {\rm Im}(I)} \,,
\end{equation}
where $\Gamma$ is the tunneling rate and ${\rm Im}(I)$ represents the imaginary part of the point-particle action. Such an imaginary contribution comes from the fact that the horizon represents a classical insuperable barrier, mathematically represented by a pole in the characteristics. In the tunneling approach, the trajectories that peel-out from the horizon are connected via a complex-path in the $x$-complex plane, which gets around the pole at the horizon. 

The link with the thermal character of the horizon is made apparent if $\Gamma$ acquires the form of a Boltzmann factor \cite{Parikh:1999mf,DiCriscienzo:2007pcr,moretti2012state,Giavoni:2020gui}
\begin{equation}
    \Gamma=e^{-\Omega/T} \,,
\end{equation}
from which the black hole temperature $T$ can be directly extracted.

Relativistic physics implies that the way null geodesics peel-out from the horizon {in stationary geometries} is universal, and in particular energy-independent. Therefore $T$ would be determined just by geometrical quantities (i.e. the horizon's surface gravity) and it would define an exactly-thermal spectrum associated to the black hole. However, in the case of modified dispersion, this feature disappears, since the motion of different particles entails an $\Omega$-dependence. As we shall see, this will be related to {a deviation from the thermal character of the emission.}

\subsection{Tunneling with modified dispersion: the approximant}
Once again, a wavepacket travels, within its geometric optic approximation, along its characteristics, defined by \cref{eq:x_characteristics}. If $k_+(x)$ enjoys a simple pole at $x=x_0$, the rate of production can be computed by connecting the two classical {solutions}  (inside and outside) through a complex path in the $x$-plane. Doing so, the action $S_\Omega(x)$ acquires an imaginary part given by
\begin{equation}
    {\rm Im}(S_\Omega)= \lim_{\epsilon \to 0} {\rm Im} \left[\int_{x_0-\epsilon}^{x_0+ \epsilon} k_+(x) {\rm d}x \right] \,.
\end{equation}

The main advantage of using the approximant as a tunneling path is the following: when modified dispersion relations are considered, the characteristics computed through the geometric optics approximation do not, strictly speaking, show any simple pole structure. The Hawking quanta outside of the horizon connects smoothly with a hard mode inside the horizon, due to its superluminal character, while the Hawking partner connects to the hard orange mode through the turning point, see Figure \ref{fig:characteristics}. 

However, if the modes are sufficiently low-energy in the sense of \cref{eq:perturbative_param}, the approximant mimics their behaviour up to regions which are very close to the EFH. One can thus wonder if, in this regime, the tunneling formula, applied to the approximant, would give the same black hole evaporation rate as computed by the usual Bogoliubov method. 

Although some heuristic justification has already been given in \cite{DelPorro:2024tuw} for expecting so, the main aim of this work is to extend the analytical Bogoliubov-based treatment of \cite{Coutant:2011in} so to include the effects of dispersion on the location of the horizon and on the surface gravity. In doing so, we shall prove the equivalence of such approach with the approximant-based tunneling calculation of \cite{DelPorro:2024tuw}.

Therefore, let us derive first what happens if we compute the tunneling rate with the approximant. From \cref{eq:S-pole}, one easily sees that the tunneling occurs at the EFH and it is ruled by $\kappa_{\rm eff}$, so that the imaginary part of the action reads
\begin{equation}
    {\rm Im}(S_\Omega)= \lim_{\epsilon \to 0} {\rm Im} \left[\int_{x_0-\epsilon}^{x_0+ \epsilon} \frac{\Omega}{\kappa_{\rm eff}(x-x_0- i \epsilon)}  {\rm d}x \right]= \frac{\pi \Omega}{\kappa_{\rm eff}} \,.
\end{equation}

The tunneling rate is therefore
\begin{equation} \label{eq:effective_rate}
    \Gamma= \exp \left[ - \frac{2 \pi \Omega}{\kappa_{\rm eff}}\right] \,,
\end{equation}
which is reminiscent of a Boltzmann distribution with temperature
\begin{equation} \label{eq:T_efh}
    T_{\rm eff}= \frac{\kappa_{\rm eff}}{2 \pi} \,.
\end{equation}

It is important to stress that this result is not describing a thermal spectrum, strictly speaking. This is obvious from the very definition of the EFH. Each frequency $\Omega$ is associated to a particular horizon's position through \cref{eq:EFH} (i.e~to a pole's location $x_0(\Omega)$) and to some $\Omega_{\rm eff}$. This fact leads to an energy dependent effective surface gravity, \cref{eq:kappa_eff}, which summarizes the expected deviations from thermality induced by the modified dispersion relation~\cite{Isoard:2019buh,DelPorro:2024tuw}.

The main point of the next Section will be to prove that, if the Bogoliubov coefficient formalism of \cite{Coutant:2011in} is applied to our set up, the evaporation's rate will match exactly the one found with the tunneling approach.

\section{Particle production via Bogoliubov coefficients: the S-matrix}
\label{sec:s-matrix}
An alternative way to express particle production by black holes is through computing the Bogoliubov coefficients, between the near-horizon modes and the ones in the asymptotic regions. These coefficients encode both the thermal character of the black hole's emission and the gray-body factor due to their propagation. Given their complexity, their evaluation is usually achieved through numerical methods (see e.g.~\cite{Macher:2009nz, Finazzi:2010yq, Finazzi:2011jd}). However, in \cite{Coutant:2011in}, the authors provide an analytical treatment to compute them within an $S$-matrix approach, in the low-energy regime.

Schematically, the region around the horizon acts on the modes as a scattering matrix $U_{\rm BH}$. The \textit{in}-state is represented by the modes moving towards the horizon (on any side), while the \textit{out}-state is given by the ones departing from it (on any side). Within low-energy regimes, i.e. $\Omega \ll \Lambda$, one can show that the ingoing mode essentially decouples (see e.g. \cite{Macher:2009tw}) and the black hole's scattering matrix becomes effectively a $2 \times 2$ matrix connecting the asymptotically incoming modes $\varphi_\Omega^{\rm or}$, $\varphi_\Omega^{\rm red}$ (corresponding to the ``hard" branches of our orange and red modes) with the asymptotically outgoing ones $\varphi_\Omega^{\rm H}$, $\varphi_\Omega^{\rm P}$ (corresponding to the ``soft" branches of our orange and red modes).

The probability conservation requires $U_{\rm BH} \in U(1,1)$, so that
\begin{align} 
    \begin{pmatrix}
        \varphi_\Omega^{\rm H}  \\[4pt]
        \varphi_\Omega^{\rm P}
    \end{pmatrix} =U_{\rm BH} \begin{pmatrix}
        \varphi_\Omega^{\rm red} \\[4pt]
        \varphi_\Omega^{\rm or}
    \end{pmatrix}=\begin{pmatrix}
        \alpha_\Omega & \beta _\Omega \\[4pt]
       \tilde \beta_\Omega & \tilde \alpha _\Omega
    \end{pmatrix}\begin{pmatrix}
        \varphi_\Omega^{\rm red} \\[4pt]
        \varphi_\Omega^{\rm or}
    \end{pmatrix}\,,
\end{align}
with unit determinant
\begin{align} \label{eq:completeness}
   |\alpha_\Omega|^2-|\beta_\Omega|^2=1\,.
\end{align}
This picture can be interpreted as a process happening along the time coordinate $t$, where the two modes $(\varphi_\Omega^{\rm red},\, \varphi_\Omega^{\rm or})$ living in the past are mapped in the future into a combination of $(\varphi_\Omega^{\rm H},\, \varphi_\Omega^{\rm P})$. The entries of $U_{\rm BH}$ are the Bogoliubov coefficients and the minus sign in the completeness relation \eqref{eq:completeness} emphasizes the presence of negative norm modes.

In \cite{Coutant:2011in}, the authors propose a different view of the same process. Instead of regard it as a past-future scattering along $t$, the static nature of the geometry at hand allows to perform the calculation of the Bogoliubov coefficients through the \qu{transfer matrix} $\mathcal U\in SL(3, \mathbb{C})$, which maps the modes \textit{inside the horizon} into the ones \textit{outside the horizon}. As in the previous picture, the coefficients $\alpha_\Omega$ and $\beta_\Omega$ appear as entries in the matrix. In this section, we will show how to derive these coefficients within this picture, providing a generalization of the treatment of \cite{Coutant:2011in} beyond the leading order in $\Omega/\Lambda$.

Technically, the advantage of this last approach is to rely on a \qu{connection formula}, where the solutions of \cref{eq:MDR_combined} in the near-horizon region are expressed as complex-contour integrals in the space of spatial momenta $k$, that directly allows to link linear combinations of modes inside and outside the horizon. Being the key ingredient, let us revise how this method works and how it can be adapted to go beyond the low energy regime.

\subsection{\textit{k}-representation}
The mode analysis of Sect.\ref{subsec:modes} describes the WKB solutions of \cref{eq:modified_KG} in the asymptotic regions. However, such an approximation fails while approaching the horizon, so to give rise to  particle production phenomena. As explained in~\cite{Coutant:2011in}, instead of their $x
$-WKB expression, one should work in the $k$-representation where the modes $\phi_\Omega(x)$ are expressed as 
\begin{equation} \label{eq:contour}
    \phi_\Omega (x)= \int_{\mathcal C} \frac{{\rm d}k}{\sqrt{2\pi}} \Tilde{\phi}_\Omega(k) \,  e^{ikx} \,,
\end{equation}
where $\mathcal C$ is a contour in the complex $k-$plane. Note that \cref{eq:contour} is the standard Fourier transform only when $\mathcal C = \mathbb{R}$, while, in general, different choices of $\mathcal C$ correspond to different solutions of \cref{eq:modified_KG}. In particular, we shall see how choosing contours which are \textit{not} homotope to the real line allow us to construct independent solutions to complete our basis.

The key point of \cite{Coutant:2011in} is that, while the $x$-WKB breaks down near the horizon due to particle production (mode-mixing), the shape of the modes in the $k$-space is well described by the $k$-WKB approximation, as long as dispersive effects can be treated perturbatively, namely $k \ll \Lambda$. The relation of \cref{eq:MDR_combined} can be written in terms of the variable $k$ as
\begin{equation} \label{eq:characteristics}
    \Omega - v \left( X_\Omega (k) \right) k=  \pm F(k)
\end{equation}
where the position $X_\Omega(k)$ is now a function of $k$. The point particle action $I=- \Omega t + S_\Omega(x)$, with
\begin{equation}
   S_\Omega(x)= \int^x k_\Omega(x') \, {\rm d } x' \,,
\end{equation}
can be written in the $k$-space by Legendre-transforming $S_\Omega(x)$
\begin{equation}
   W_\Omega(k):=-kx + S_\Omega(x)=: -\int^k X_\Omega(k') \, {\rm d } k'\,.
\end{equation}
The $k$-WKB mode then takes the form \cite{Coutant:2011in}
\begin{equation}
   \Tilde{\varphi}_\Omega(k)= \sqrt{\frac{\partial X_\Omega(k)}{\partial \Omega}} \frac{e^{i W_\Omega (k)}}{\sqrt{4 \pi F(k)}}\,.
\end{equation}
Comparing with \cref{eq:contour}, one can immediately see that the mathematical root for particle production is hidden in the analytic properties of $W_\Omega(k)$. As a function of the complex variable $k$, the presence of poles and branch-cuts will denote mode-mixing mechanisms, as we shall see in a moment.

\subsection{Analytic properties of $W_\Omega (k)$}
The determination of the analytic structure of $W_\Omega (k)$ is crucial for computing the Bogoliubov coefficients. In \cite{Coutant:2011in}, such an analysis has been performed working in the low-energy regime, where $W_\Omega (k)$ is shown to enjoy a branch-cut in the $k$-complex plane whenever $x \simeq 0$. That branch-cut is associated to the sonic regime of the dispersion relation and to the particle production from the acoustic horizon. Here, we want to show how to extend such a treatment beyond the sonic regime. 

The starting point is based on the following consideration: as noted in \cite{Jacobson:2003vx}, the mode mixing process, which is at the basis of Hawking radiation, can be determined by just looking as how the different outgoing modes (both the soft and the hard branches) interact with each other. 

{As already mentioned,} in principle also the ingoing mode is found to interact with the outgoing branches, contributing to the gray-body factor, and so to the resulting spectrum at infinity. However, its contribution (especially in the regime where $k \ll \Lambda$, for which massless modes in $(1+1)$ dimensions effectively decouple) is usually negligible with respect to the interaction between the outgoing branches. So, following \cite{Coutant:2011in}, here we shall neglect its contribution and treat it as a spectator mode.

Nontheless, as shown in \cite{DelPorro:2024tuw}, such a mode appears to be extremely important when coming to define the approximant trajectory. Let us show how this is realized in the formalism of \cite{Coutant:2011in} and how this connects to the analytic properties of $W_\Omega (k)$ beyond the pure sonic regime.

\subsubsection{Outgoing mode outside the EFH}
Following Section \ref{subsec:approximant}, we start by considering the Hamilton--Jacobi equations for the ingoing and the (soft red) outgoing modes outside of the acoustic black hole
\begin{subequations}
    \begin{align}
    &\Omega - v \left( X^+_\Omega (k) \right) k= F(k)  \qquad \mbox{outgoing}\,, \label{eq:outgoing}\\
    &\Omega - v \left( X^-_\Omega (k) \right) k= F(k) \qquad \mbox{ingoing} \,, \label{eq:ingoing}
\end{align}
\end{subequations}
where $X_\Omega^\pm$ represents the position of the ingoing ($-$) and outgoing ($+$) mode in the $k$-representation.

We are interested in computing the $k$-dependence of the outgoing mode $X_\Omega^+(k)$. To do so, we operate as in the previous sections: since the ingoing mode experiences the whole spacetime in a continuous way, one can keep track of the outgoing trajectory $X_\Omega^+$ making use of the ingoing one $X^-_\Omega$. In other words, given an ingoing mode with some wavenumber $k_-$ and position $X^-_\Omega(k_-)$, we \textit{define} $k_+=k_+(k_-)$ as the quantity satisfying
\begin{equation} \label{eq:equal_positions}
    X^+_\Omega(k_+)= X^-_\Omega(k_-) \,.
\end{equation}
Note that, in the $x$-representation, this corresponds to solve both equation at fixed $x$, as done above. In particular, one can use the same parametrisation as in \cref{eq:x_parametrization} and solve for $R(k_-)$. However, this time $k_-$ is the independent variable instead of $x$.

The steps operated in the $x$-representation were algebraic manipulations, which can be repeated in the same way in the $k$-picture. What changes here is that now the definition of the EFH becomes an equation for $k_-$
\begin{align} \label{eq:k_EFH}
    v \left(X^-_\Omega(k_-) \right)-c_g(k_-) =0\,.
\end{align}
Once again, if $\bar k$ is a solution of that equation (so that $X_\Omega^-(\bar k) := x_0$) and \cref{eq:perturbative_param} holds, we get
\begin{equation} \label{eq:k_approximant}
    R(k_-)= \frac{2 v \left( X_\Omega^- \right)  k_-}{v \left( X_\Omega^- \right) -c_g^- }\,.
\end{equation}
At the next-to leading order, one can take into account perturbatively the role of $ \left(4C_2 \bar v \bar k/\Lambda \right)^{\gamma} $ in computing the Bogoliubov coefficient. We shall return to this point later.

Eq.\eqref{eq:k_approximant} can be used to compute the effective action $W_\Omega(k_+)$ as
\begin{align} \label{eq:chain_rule}
    W_\Omega(k)= -\int^{k} X^+_\Omega(k_+) {\rm d}k_+ =-\int^{f^{-1}(k)} X^-_\Omega(k_-) \frac{{\rm d}k_+}{{\rm d}k_-}{\rm d}k_-
\end{align}
with
\begin{equation}
     \frac{{\rm d}k_+}{{\rm d}k_-}=-1+ R'(k_-) \,.
\end{equation}

As we showed above, $R(k_-)$ is continuous except for the region close to the solution of \cref{eq:k_EFH}, where it shows a simple pole. Indeed, from \cref{eq:k_approximant}, we have (the argument of the functions is understood)
\begin{equation}
      R'(k_-)= \frac{\partial_{k_-}(2  v k_-)}{v-c_g^-}  - \frac{\partial_{k_-}(v-c_g^-)(2  v k_-)}{(v-c_g^-)^2}  \,.
\end{equation}
and, for $\delta k=k_- - \bar k \simeq 0$ we get that $R '(k_-)$ contains an irregular part
\begin{equation}
      R'{}^{\rm irr}(k_-)=\frac{2  \bar v \bar k}{\partial_{k_-}(v-c_g^-)}\biggl|_{\bar k} \left[ \frac{\partial^2_{k_-}(v-c_g^-)}{\partial_{k_-}(v-c_g^-)}\biggl|_{\bar k}\frac{1}{\delta k} - \frac{1}{\delta k^2} \right]  \,.
\end{equation}
At the same time, one can expand $X^-_\Omega(k_-)$ around $x_0=X^-_\Omega(\bar k)$
\begin{equation}
     X^-_\Omega(k_-)=x_0+ \frac{{\rm d}X^-_\Omega}{{\rm d}k_-} \biggr|_{\bar k} \delta k + \frac 12 \frac{{\rm d}^2X^-_\Omega}{{\rm d}k_-^2} \biggr|_{\bar k} \delta k^2 \cdots\,.
     \label{eq:Xexpans}
\end{equation}
Some comments are in order. First, we can see that near $\bar k$, the action $W_\Omega(k)$ presents some non-analiticities: 
\begin{align}
\label{eq:Wirr}
     W_\Omega^{\rm irr}(k)=-\frac{2  \bar v \bar k x_0}{\partial_{k_-}(v-c_g^-)} \biggl|_{\bar k} \frac{1}{\delta k}-\frac{2  \bar v \bar k}{\partial_{k_-}(v-c_g^-)}\biggl|_{\bar k} \left[ x_0\frac{\partial^2_{k_-}(v-c_g^-)}{\partial_{k_-}(v-c_g^-)} - \frac{{\rm d}X^-_\Omega}{{\rm d}k_-}  \right]_{\bar k} \ln(\delta k)  \,,
\end{align}
where here $\delta k =k_-(k)- \bar k$ should be read with $k_-(k)=f^{-1}(k)$ as the inverse of \cref{eq:x_parametrization}.
These non analiticities will be the source of the particle production from the EFH located at $x=x_0$. Here, we notice that the role of the surface gravity in the $k$-representation is played by the $k$-derivative of the function $(v-c_g^-)$. Analogously to what usually done in the $x$-representation, one can define an \textit{adiabaticity condition} for the mode in the $k$-representation to be well described by its WKB form. 

In particular, if we ask the first term in the square bracket of \cref{eq:Wirr} to be negligible w.r.t.~the second one, we have to require 
\begin{equation}
\label{eq:k_adiabaticity}
  \frac{\partial^2_{k_-}(v-c_g^-)}{(\partial_{k_-}(v-c_g^-))^2}\biggr|_{\bar k} \ll \frac{1}{\tilde\kappa_\textsc{efh} x_0} \,,
\end{equation}
where
\begin{equation} \label{eq:k_kappa}
    \frac{1}{\tilde\kappa_\textsc{efh}}:= \frac{{\rm d}X^-_\Omega}{{\rm d}k_-}  \frac{1}{\partial_{k_-}(v-c_g^-)} \biggl|_{\bar k}= \frac{1}{\partial_{X_\Omega^-}(v-c_g^-)} \biggl|_{\bar k}\,.
\end{equation}
which has the physical interpretation to require the variation of the surface gravity to be adiabatic in the momentum near the EFH at $k_-= \bar k$.

We emphasize that this $k$-adiabatic regime always holds for relativistic dispersion relations. In that case, $F(k^2)^2=k^2$ and the EFH coincides with the Killing horizon, thus $x_0=0$, rendering \cref{eq:k_adiabaticity} exact.

Additionally, looking at the expansion \eqref{eq:Xexpans} and calculating it in $\delta k =0$, one can clearly see that the definition of $\tilde \kappa_{\textsc{efh}}$ matches the peeling surface gravity of \cref{eq:Xkappa} and of \cite{DelPorro:2024tuw}. For this reason we shall omit from now on the tilted notation in spite of the different definition of the ``$k$-representation" surface gravity with respect to the standard one \cref{eq:Xkappa}. Note however, that only in the adopted adiabatic regime $\tilde{\kappa}_\textsc{efh}$ is the single term determining the coefficient in front of the logarithmic term of \cref{eq:Wirr}. From now on such adiabatic regime will be implicitly assumed.

\subsubsection{Outgoing mode inside the EFH}
A similar reasoning can be done for the soft outgoing mode inside the EFH. In this case we should consider
\begin{subequations}
    \begin{align}
    -\Omega - v \left( X^+_\Omega (k) \right) k= -F(k)  \qquad &\mbox{outgoing}\,, \label{eq:outgoing_inside}\\
    \Omega - v \left( X^-_\Omega (k) \right) k= F(k) \qquad &\mbox{ingoing} \,, \label{eq:ingoing_inside}
\end{align}
\end{subequations}
and an analogous parametrization
\begin{align}
\label{eq:k+}
    k_+(k_-)=k_-+R(k_-) \,.
\end{align}
Within the same approximations of the previous section we get
\begin{equation} \label{eq:r_k-_2}
  R(k_-) = -\frac{2 v \left( X_\Omega^- \right)  k_-}{v \left( X_\Omega^- \right) -c_g^- } \,.
\end{equation}
Here, we can recognize the same structure of the outgoing mode outside the horizon, but with a reversed sign for $R(k_-)$. That implies the irregular part of $W_\Omega(k)$ acquiring a minus sign in front of the simple pole and the logarithm to be function of $\bar k- k=-\delta k$
\begin{align}
     W_\Omega^{\rm irr}(k)=\frac{2  \bar v \bar k x_0}{\partial_{k_-}(v-c_g^-)} \biggl|_{\bar k} \frac{1}{\delta k} -\frac{2  \bar v \bar k}{\partial_{k_-}(v-c_g^-)}\biggl|_{\bar k} \left[ x_0\frac{\partial^2_{k_-}(v-c_g^-)}{\partial_{k_-}(v-c_g^-)} - \frac{{\rm d}X^-_\Omega}{{\rm d}k_-}  \right]_{\bar k} \ln(-\delta k) \nonumber \,.
\end{align}

\subsection{The connection formula and the S-matrix}
Recall the formula
\begin{align} \label{eq:chain_rule_2}
    W_\Omega = -\int^{f^{-1}(k)} X^-_\Omega(k_-) \frac{{\rm d}k_+}{{\rm d}k_-}{\rm d}k_- = -x_0 k_+(k_-) + \frac{2 \bar v \bar k}{\kappa_\textsc{efh}} \ln(\delta k) + \cdots.
\end{align}

It is convenient to locally invert the function $k_+(k_-)$ as given in \cref{eq:k+} and \cref{eq:r_k-_2}, around $\bar k$. We get
\begin{equation} \label{eq:inverse}
    k_+ \simeq \frac{2 \bar v \bar k}{\kappa_\textsc{efh} \delta k}\left.\frac{{\rm d}X^-_\Omega}{{\rm d}k_-}\right|_{\bar{k}} \iff \delta k \simeq \frac{2 \bar v \bar k} {\kappa_\textsc{efh} k_+}\left.\frac{{\rm d}X^-_\Omega}{{\rm d}k_-}\right|_{\bar{k}}\,,
\end{equation}
from which we get, together with \cref{eq:Omega_eff} and \cref{eq:kappa_eff}
\begin{equation} 
    W_\Omega= - x_0 k_+ - \frac{\Omega}{\kappa_{\rm eff}} \ln(k_+) + \mathcal O(1).
\end{equation}

The integral to compute is then  
\begin{equation} \label{eq:reference_integral}
   \phi_\Omega(x)= \int_{\mathcal C} \frac{{\rm d} k}{\sqrt{2 \pi}}\sqrt{\frac{\partial X^+_\Omega}{\partial \Omega}} \frac{e^{ikx} e^{iW_\Omega(k)}}{\sqrt{4 \pi  F(k,\Lambda)}} 
\end{equation}
which, near the EFH $k_-= \bar k$ takes the form
\begin{equation} \label{eq:near_EFH}
   \phi_\Omega(x)= \int_{\mathcal C} \frac{{\rm d}k}{\sqrt{2 \pi}} \sqrt{\frac{\partial X^+_\Omega}{\partial \Omega}}\frac{e^{ik(x-x_0)} e^{ -i(\Omega/\kappa_{\rm eff})\ln(k)}}{\sqrt{4 \pi F(k)}} \,.
\end{equation}

Therefore, in the complex $k$-plane, the action $W_\Omega$ exhibits a branch cut for $k=0$ as it happens in \cite{Coutant:2011in}. Choosing the position of such a branch-cut appropriately, together with the integration contour $\mathcal C$, gives the possibility to compute the $3 \times 3$ scattering matrix between the modes inside and the modes outside the acoustic black hole. 

The way $\mathcal{C}$ goes at infinity (for $|k| \to \infty$) can be determined by looking at the behaviour of the phase in \cref{eq:reference_integral}. In practice, integrating by parts we get
\begin{equation}
    W_\Omega(k)=\int^k X^+_\Omega(k_+) dk_+=X^+_\Omega(k) k+ \int^k \frac{c_g^+(k_+)+v}{v'} {\rm d}k_+ \,.
\end{equation}
Following \cite{Coutant:2011in} let us introduce the rescaling $k=\Lambda s$ where $s$ is a complex parameter (note that in \cite{Coutant:2011in} $s=t$). We have
\begin{align}
    W_\Omega(k)&=\Lambda X^+_\Omega(\Lambda s) s + \int^{\Lambda s} \frac{c_g^+(k_+)+v}{v'} {\rm d}k_+ \simeq \Lambda X^+_\Omega(\Lambda s) s + \mbox{sgn}(s) \Lambda\frac{|s|^n}{n v'}\,.
\end{align}
where in the last step we take into account the leading order in $|s|\to \infty$ in the integral and the sign depends on the sign of the group velocity. Therefore, at infinity, the contour $\mathcal{C}$ must acquire, in the $s$-complex plane, a small positive imaginary part. That is, the contour goes from $-\infty + i \varepsilon$ to $+ \infty + i\varepsilon$, with an angle $\theta_{\pm \infty}$ within the range
\begin{equation} \label{eq:theta_infty}
0< \theta_{+\infty}< \frac{\pi}{n} \,, \qquad \frac{n-1}{n} \pi< \theta_{-\infty}< \pi \,.
\end{equation}
In particular, this implies that the region just below the real line (i.e. $\pm \infty - i \epsilon$) represents a sort of \qu{forbidden region} where to close the contour. 

Additionally, we see that when $\Lambda \to \infty$, the phase acquires a linear dependence in $\Lambda$, so that we can apply the saddle point approximation
\begin{equation} \label{eq:steepest_descent}
\int_C A(s) e^{i\Lambda h(s)} \simeq A(s_j)\frac{e^{i \Lambda h(s_j)}}{\sqrt{-i h''(s_j)}}\left(1+ \mathcal O\left(\frac{E_j}{\Lambda} \right) \right)\,.
\end{equation}
with $h(s)$ being an arbitrary function and $s_j$ satisfying
\begin{equation}
\frac{{\rm d}h(s)}{{\rm d}s}\biggr|_{s_j}=0\,,
\end{equation}
and $E_j$ depending on second and higher derivatives of $h$ and $A$ computed at the saddle point $s_j$ \cite{olver2014asymptotics,Coutant:2011in}.

We emphasize that the saddle point of the phase in the integrand of \cref{eq:reference_integral} corresponds (up to a global phase) to the solution in the $x$-representation. Indeed if $k_j=\Lambda s_j$ is a saddle point, it has to respect
\begin{equation} \label{eq:saddle_point}
0=\frac{{\rm d}}{{\rm d}k}\left(kx- \int^kX^+_\Omega(k_+) {\rm d}k_+ \right) \implies x=X^+_\Omega(k_j)\,,
\end{equation}
which precisely identifies the different outgoing branches of the solution in the $x$-representation.

\subsection{The decaying mode}
\begin{figure}[t!]
    \centering
    \includegraphics[scale=0.16]{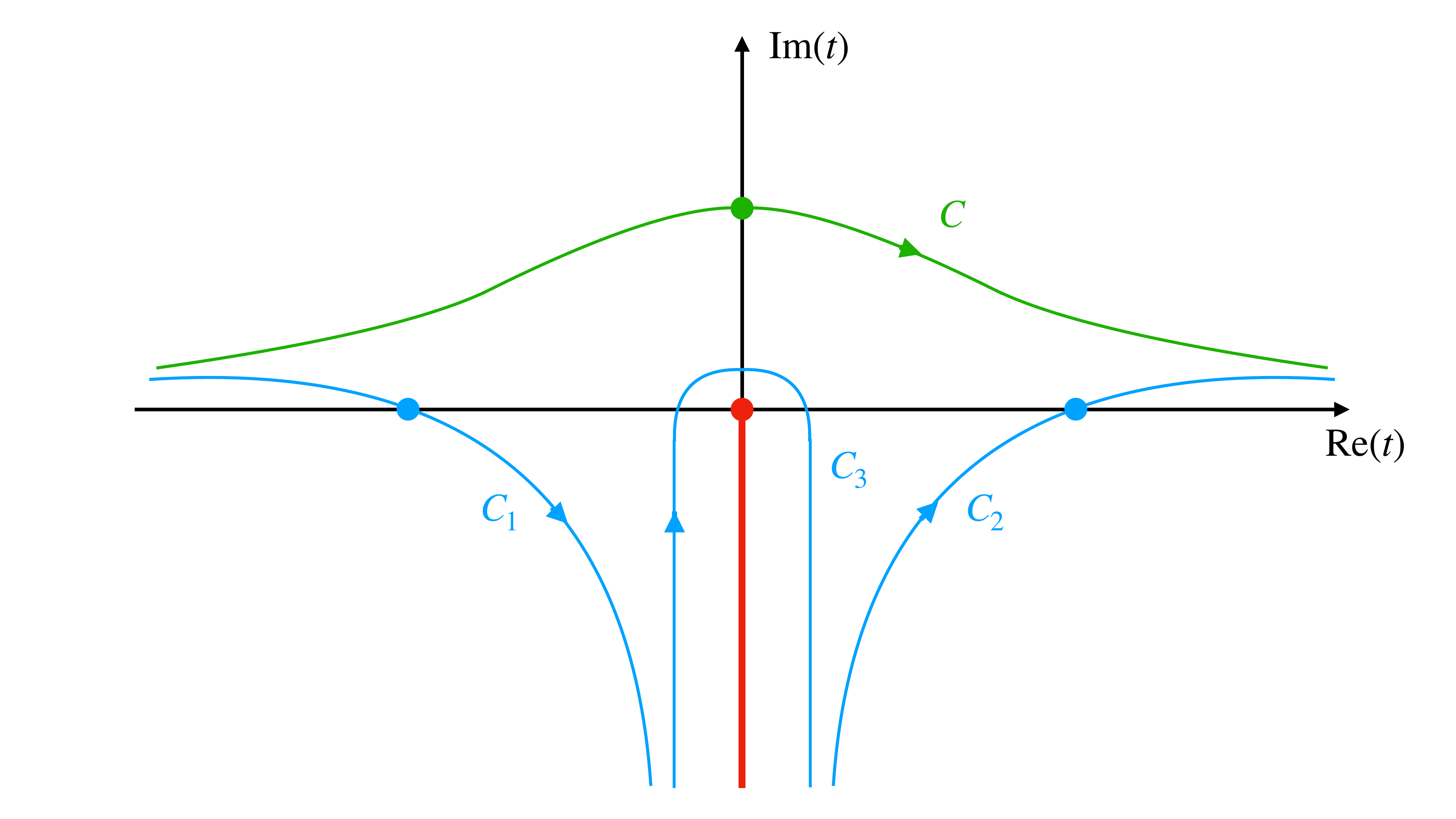}
    \caption{Contours of integration for the decaying mode. The light blue line (that corresponds to the case $x<x_0$) is composed by $\mathcal{C}_1$, $\mathcal{C}_2$ and $\mathcal{C}_3$. It gets around the branch cut -- the thick red line along the negative imaginary axis -- and crosses the two real saddle points (light blue dots). The green line corresponds to the contours for $x>x_0$ and crosses the imaginary saddle point (green dot), i.e. the decaying mode.} 
    \label{fig:Contour_decaying}
\end{figure}

Following closely the treatment of \cite{Coutant:2011in}, we start by placing the branch-cut along the negative imaginary axis $\{k=i y; \, y<0 \}$, as depicted in Figure \ref{fig:Contour_decaying}.

For $x<x_0$, the contour can be chosen to start from $k=-\infty + i \epsilon$ and split in three different contributions:
\begin{itemize}
    \item $\mathcal{C}_1$: crosses the real axis up to $k=-\epsilon -i \infty$,
    \item $\mathcal{C}_3$: gets around the branch cut from $k=-\epsilon -i \infty$ to $k=\epsilon -i \infty$,
    \item $\mathcal{C}_2$: starts from $k=\epsilon -i \infty$, crosses the real axis and ends in $k=\infty + i \epsilon$.
\end{itemize}

The total integral along $\mathcal C= \mathcal{C}_1 \cup \mathcal{C}_2 \cup \mathcal{C}_3$ can be divided into a sum of three integrals. Two of them, $\mathcal{C}_1 $ and  $\mathcal{C}_2 $, can be computed at first order in $\Omega/\Lambda$ through the saddle point approximation. The solution of \cref{eq:saddle_point} is given by $k_\Lambda^{\rm or}(x)<0$. To evaluate the different contribution to the phase given by the square root in \cref{eq:steepest_descent}, we need to compute the sign of
\begin{align}
    \frac{{\rm d}^2}{{\rm d}k^2} \left( kx- \int^kX^+_\Omega(k_+) {\rm d}k_+ \right)_{k_\Lambda^{\rm or}}= \frac{c_g(k)+v(x)}{k v'(x)} \biggr|_{k_\Lambda^{\rm or}} \,.
\end{align}
We point out that, from the Hamilton equation $c_g(k(x))+v(x)= $d$x/$d$t>0$, since we are considering the hard modes traveling towards the horizon from inside. Additionally $v'(x)>0$, therefore 
\begin{align}
   \mbox{sgn}\left( \frac{c_g(k)+v(x)}{k v'(x)} \biggr|_{k_\Lambda^{\rm or}}\right)= \mbox{sgn}\left( k_\Lambda^{\rm or}\right)<0\,.
\end{align}
In this way, from \cref{eq:steepest_descent}, the saddle acquires a phase of $\sqrt{-i}=e^{3i \pi /4}$, namely
\begin{align}
   \varphi_\Omega^{\mathcal C_1}=e^{3i \pi /4} \varphi_\Omega^{\rm or}\,.
\end{align}

The discussion on $\mathcal C_2$ is analogous, with the saddle point being $k_\Lambda^{\rm red}>0$, so that the mode acquires a phase of $\sqrt{i}=e^{i \pi /4}$
\begin{align}
   \varphi_\Omega^{\mathcal C_2}=e^{i  \pi /4} \varphi_\Omega^{\rm red}\,.
\end{align}

The integral on $\mathcal C_3$, instead, can be computed as in \cref{eq:near_EFH}. Differentiating \cref{eq:outgoing} with respect to $\Omega$ we have:
\begin{equation}
    \frac{\partial X^+_\Omega}{\partial \Omega}= \frac{1}{v'(X^+_\Omega) k} \,,
\end{equation}
so that the integral becomes
\begin{equation} \label{eq:around_branch_cut}
    \int_{\mathcal C_3} \frac{{\rm d}k}{k}\frac{e^{ik(x-x_0)} e^{ i(\Omega/\kappa_{\rm eff})\ln(k)}}{\sqrt{8 \pi^2 v'(x_0)}} \left( 1- \frac{c_2}{4} \frac{k^2}{\Lambda^2} + \cdots \right) \,.
\end{equation}
where we Taylor-expanded $F(k)=k(1+c_2 k^2/\Lambda^2 + \cdots)$ around $k/\Lambda=0$. The leading order, given by standard integral formulas \cite{Gradshteyn:1943cpj}, is 
\begin{align} \label{eq:around_branch_cut_LO}
   \varphi_\Omega^{\mathcal C_3}\simeq\int_{\mathcal C_3} \frac{{\rm d}k}{k}\frac{e^{ik(x-x_0)} e^{ i(\Omega/\kappa_{\rm eff})\ln(k)}}{\sqrt{8 \pi^2 |v'(x_0)|}}= - \frac{(x-x_0)^{i \Omega/\kappa_{\rm eff}}}{\sqrt{2 \pi^2 |v'(x_0)|}} \sinh \left( \frac{\pi \Omega}{\kappa_{\rm eff}} \right) \Gamma \left( -i\frac{\Omega}{\kappa_{\rm eff}} \right)  e^{ \pi\Omega/ 2\kappa_{\rm eff}}\,.
\end{align}
The next-to-leading contribution of \cref{eq:around_branch_cut} can be easily obtained by differentiating twice \cref{eq:around_branch_cut_LO} by $x$, i.e.
\begin{align} \label{eq:around_branch_cut_LO_twice}
    &- \frac{c_2}{4}\int_{\mathcal C_3} \frac{{\rm d}k}{k}\frac{e^{ik(x-x_0)} e^{ i(\Omega/\kappa_{\rm eff})\ln(k)}}{\sqrt{8 \pi^2 |v'(x_0)|}} \frac{k^2}{\Lambda^2}=  \frac{c_2}{4 \Lambda^2} \frac{\partial^2}{\partial x^2}\int_{\mathcal C_3} \frac{{\rm d}k}{k}\frac{e^{ik(x-x_0)} e^{ i(\Omega/\kappa_{\rm eff})\ln(k)}}{\sqrt{8 \pi^2 |v'(x_0)|}}=  \\
   &- \frac{c_2}{4}\frac{(x-x_0)^{i \Omega/\kappa_{\rm eff}}}{\sqrt{2 \pi^2 |v'(x_0)|}} \sinh \left( \frac{ \pi\Omega}{\kappa_{\rm eff}} \right) \Gamma \left( 2-i\frac{\Omega}{\kappa_{\rm eff}} \right)   \frac{e^{ \pi \Omega /2 \kappa_{\rm eff}}}{\Lambda^2 (x-x_0)^2} \nonumber \,.
\end{align}
That means that the next-to-leading order starts to contribute within a region around the horizon of the order of the cutoff length
\begin{equation} \label{eq:NLO_branchcut_parameter}
    |x-x_0| \simeq \frac{1}{\Lambda} \,.
\end{equation}
Since $\Lambda^{-1}$ represents a cutoff length below which the system cannot be resolved, we can safely neglect such terms.\footnote{That is precisely what the authors do in \cite{Coutant:2011in}, by applying the \qu{dominated convergence theorem} along the contour $\mathcal C_3$, in order to extract the leading contribution in $k/\Lambda$ in the integrand.}.

Putting altogether, the total contour along $\mathcal C$ is given by the sum of
\begin{equation}\label{eq:decaying_inside}
   \varphi_\Omega^{\mathcal C}=e^{3i \pi/4}e^{\pi \Omega/\kappa_{\rm eff}} \varphi^{\rm or}_\Omega+e^{i  \pi/4}  \varphi_\Omega^{\rm red} +  \varphi_\Omega^{\mathcal C_3}\,,
\end{equation}
where the factor $e^{\pi \Omega/\kappa_{\rm eff}}$ comes as the analytical continuation of $W_\Omega(k)$ with the branch cut in the lower imaginary axis. In other words, in this configuration $\log(k)=\log|k| +i \pi$ for $k<0$. Note that, up to a change $(\kappa_{\rm eff} \leftrightarrow \kappa_\textsc{kh})$, the expression \eqref{eq:decaying_inside} corresponds to what found in \cite{Coutant:2011in}.

Outside the horizon, namely for $x>x_0$ the same contour can be deformed to cross the upper imaginary axis in the saddle point $k^{\rm dec}_\Lambda$. This saddle is associated to the decaying mode, since ${\rm Im}(k^{\rm dec}_\Lambda)>0$, up to a global phase $e^{i \theta_{\rm dec}}$ determined by the denominator of \cref{eq:steepest_descent}. Therefore, for $x>x_0$
\begin{equation}
   \varphi_\Omega^{\mathcal C}=e^{i \theta_{\rm dec}} \varphi_\Omega^{\rm dec}\,.
\end{equation}

\subsection{The Hawking quanta}
\begin{figure}[t!]
    \centering
    \includegraphics[scale=0.16]{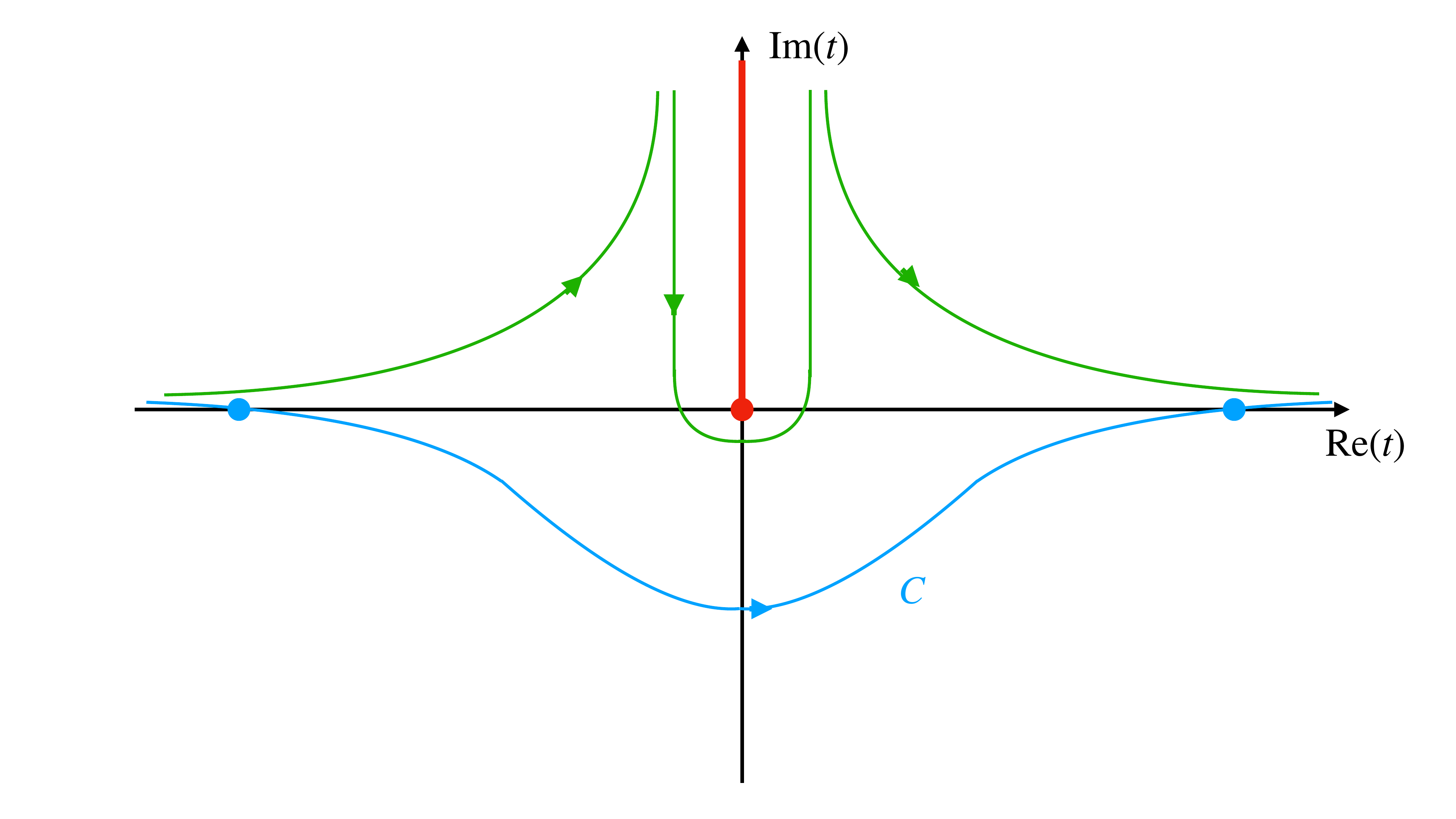}
    \caption{Contours of integration for the Hawking quanta. The light blue line again corresponds to the case $x<x_0$ and hits the two real saddles (light blue dots). The green line corresponds to the contours for $x>x_0$ get around the branch cut, which now is sitting along the positive imaginary axis.} 
    \label{fig:Hawking_quanta}
\end{figure}

Next, we put the branch cut in the upper half imaginary axis $\{k=i y; \, y>0 \}$, as shown in Figure \ref{fig:Hawking_quanta}. 
Following the same logic, we can start with $x<x_0$. The contour $\mathcal C$ crosses twice the real axis in the two saddles of the previous paragraph, but this time there is no branch cut on the way. Given the different location of the cut, the continuation of the logarithm is $\log(k) \to \log|k| - i \pi$ and gives rise to
\begin{equation}
    \varphi_\Omega^{\mathcal C}=e^{i \pi/4}\varphi_\Omega^{\rm red}+e^{3i \pi/4}e^{-\pi \Omega/\kappa_{\rm eff}}\varphi_\Omega^{\rm or} \,.
\end{equation}

For $x>x_0$ the contour is deformed to go up until $k= - \epsilon + i \infty$ and circumvent the cut up to $k=  \epsilon + i \infty$. A similar discussion with respect to the previous case leads to
\begin{equation} \label{eq:around_branch_cut_upper}
    \varphi^{\mathcal C}_\Omega= \frac{(x_0-x)^{i \Omega/\kappa_{\rm eff}}}{\sqrt{2 \pi^2 |v'(x_0)|}} \sinh \left( \frac{ \pi\Omega}{\kappa_{\rm eff}} \right) \Gamma \left( -i\frac{\Omega}{\kappa_{\rm eff}} \right)  e^{- \pi \Omega / 2\kappa_{\rm eff}}\,.
\end{equation}

\subsection{The growing mode}
The last, independent mode, would be the growing mode. As in \cite{Coutant:2011in}, we place again the branch cut along the positive imaginary axis, but this time we choose a contour of integration which cannot be continuously deformed into the real axis. As we will see, this automatically guarantees linear independence. In particular, since the branch cut lies on the upper half complex plane, we decide to perform two independent integrals. The first one is along the contour which goes from $k=- \infty+i \epsilon$ to $k= \epsilon-i \infty$ and we will name it $\tilde{\mathcal C}_1$. Note that this contour crosses the imaginary axis. The second one, named $\tilde{\mathcal C}_2$ is mirrored: it starts from $k=- \epsilon-i \infty$ and it goes up to $k=  \infty+ i \epsilon$. 

\begin{figure}[t!]
    \centering
    \includegraphics[scale=0.16]{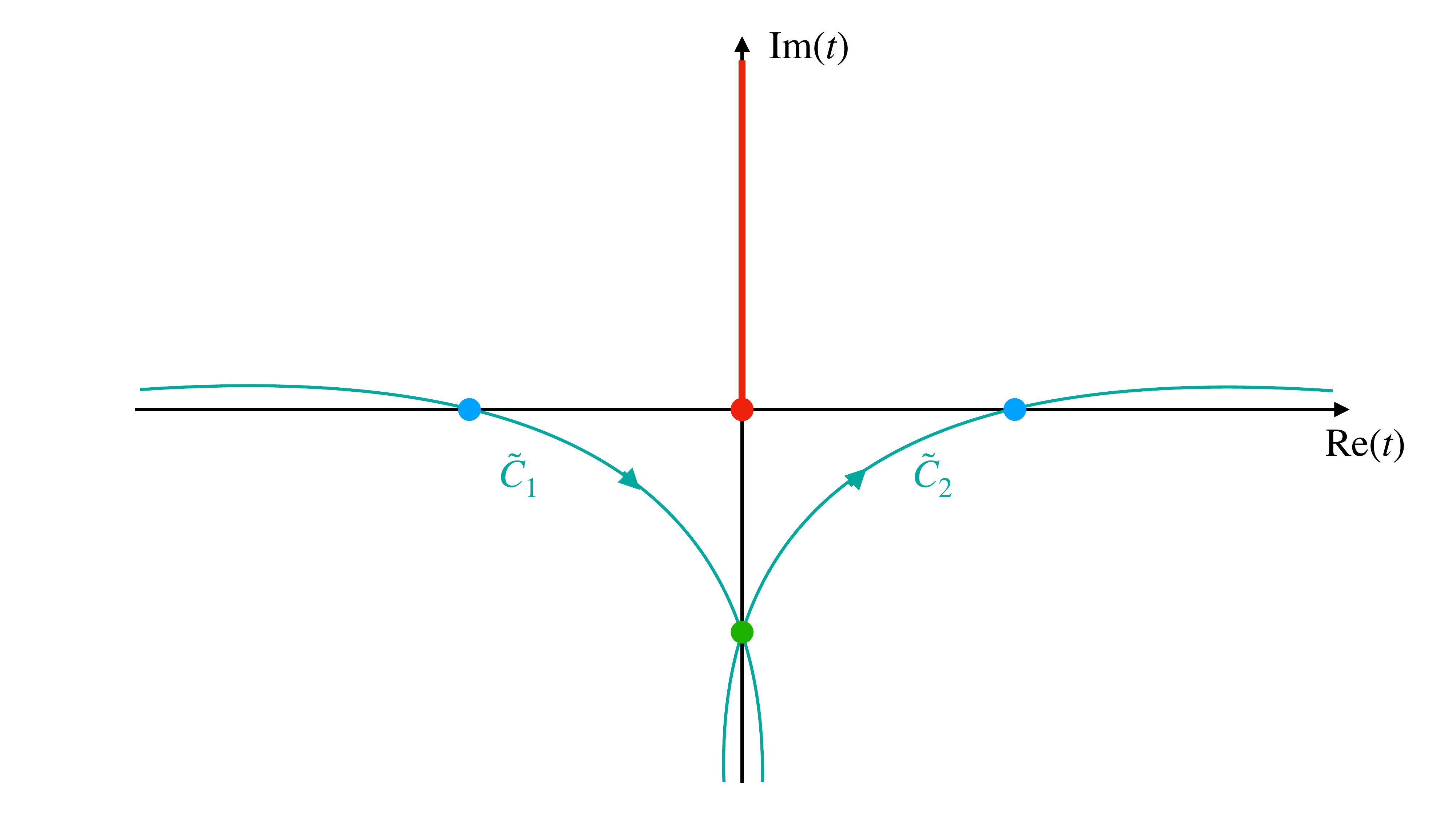}
    \caption{Contours of integration for the growing mode. Both $\tilde{\mathcal C}_1$ and $\tilde{\mathcal C}_2$ are not homotope to the real line. For $x>x_0$, both of them hit the negative imaginary saddle (green dot, corresponding to the growing mode) while for $x<x_0$ each of them hits a different real saddle point (light blue dots).} 
    \label{fig:growing}
\end{figure}

It is important that it is not possible to continuously deform these contours into the real axis, since any continuous deformation would imply to cross a region where the contour should close at $k=\pm i \infty - i \epsilon$ which is forbidden by \cref{eq:theta_infty}. Therefore the contours considered here give rise to an actual independent solution with respect to the ones studied previously in this section. The contours are represented in Figure \ref{fig:growing}. 

The versatility of these two contours is that, for $x>x_0$, they share the same saddle point along the negative imaginary axis $k=k_\Lambda^{\rm gr}$. For $x<x_0$, instead, $\tilde{\mathcal C}_1$ can be computed at the leading order with its real saddle point $k_\Lambda^{\rm or}$ and similarly  $\tilde{\mathcal C}_2$ with $k_\Lambda^{\rm red}$. At the practical level, we have:
\begin{equation} \label{eq:Ctilde_VS_C}
     \varphi_\Omega^{\tilde{\mathcal C}_1}= -e^{-2 \pi \Omega/\kappa_{\rm eff}} \varphi_\Omega^{\mathcal C_1} \,, \qquad  \varphi_\Omega^{\tilde{\mathcal C}_2}= \varphi_\Omega^{\mathcal C_2}\,.
\end{equation}
Here, the exponential correspond to continuing again the logarithm as $\ln(k) \to \ln|k| - i \pi$, due to the position of the branch cut. Following the same reasoning of \cite{Coutant:2011in}, we notice that the minus sign correspond to choosing $\sqrt{-i}=e^{-i \pi /4}$, so to give a true linearly independent solution (as we shall see later).

In this way, we have two independent ways to write the growing mode $\varphi_\Omega^{\rm gr}$ associated to $k_\Lambda^{\rm gr}$. We thus have:
\begin{equation} \label{eq:growing}
    \varphi_\Omega^{\rm gr}=  \varphi_\Omega^{\mathcal C_2}=-e^{-2 \pi  \Omega/\kappa_{\rm eff}} \varphi_\Omega^{\mathcal C_1} \,.
\end{equation}
Putting all together we get, as in \cite{Coutant:2011in}
\begin{equation} \label{eq:2growing}
     2\varphi_\Omega^{\rm gr}= e^{-i \pi /4} e^{-   \pi \Omega/\kappa_{\rm eff}} \varphi^{\rm or}_\Omega + e^{i \pi /4}  \varphi^{\rm red}_\Omega\,.
\end{equation}
Now we are ready to write down our transfer matrix.

\subsection{Transfer Matrix}
To build the transfer matrix, we need to connect the modes inside the horizon with the ones outside the horizon. This can be done easily by noticing that \cref{eq:around_branch_cut_LO} can be rewritten, with the definition given in \cref{eq:P_mode}, as 
\begin{align} \label{eq:C3_partner}
\varphi_\Omega^{\mathcal C_3}(x)=-\varphi_\Omega^{\rm P}(x) \times \mathcal{A}(\Omega)
\end{align}
where we have introduced
\begin{align}
 \mathcal{A}(\Omega)=\sqrt{\frac{2 \Omega}{ \pi |v'(x_0)|}} \sinh \left( \frac{\pi \Omega}{\kappa_{\rm eff}} \right) \Gamma \left( -i\frac{\Omega}{\kappa_{\rm eff}} \right)  e^{\Omega \pi/2 \kappa_{\rm eff}} \,.
\end{align}

Similarly for \cref{eq:around_branch_cut_upper} and \cref{eq:H_mode} we get
\begin{align} \label{eq:C_Hawking}
\varphi_\Omega^{\mathcal C}(x)=\varphi_\Omega^{\rm H}(x) \times \mathcal{B}(\Omega)
\end{align}
where
\begin{align}
 \mathcal{B}(\Omega)=\sqrt{\frac{2 \Omega}{ \pi |v'(x_0)|}} \sinh \left( \frac{\pi \Omega}{\kappa_{\rm eff}} \right) \Gamma \left( -i\frac{\Omega}{\kappa_{\rm eff}} \right)  e^{-\Omega \pi/2 \kappa_{\rm eff}} \,.
\end{align}
Therefore, we have the transfer matrix, defined as $\mathcal U \in SL(3, \mathbb{C})$ through the map \cite{Coutant:2011in}
\begin{align} 
    \begin{pmatrix}
        \varphi_\Omega^{\rm H}  \\[4pt]
        \varphi_\Omega^{\rm dec} \\[4pt]
        \varphi_\Omega^{\rm gr}
    \end{pmatrix} =\mathcal U^T \begin{pmatrix}
        \varphi_\Omega^{\rm P}  \\[4pt]
        \varphi_\Omega^{\rm red} \\[4pt]
        \varphi_\Omega^{\rm or}
    \end{pmatrix}\,.
\end{align}
From the sections above one can immediately deduce the entries of the matrix, getting
\begin{align} 
    \mathcal U=\begin{pmatrix}
        0 & - \mathcal A e^{-i \theta_{\rm dec}} & 0  \\[4pt]
          \frac{e^{i  \frac{\pi}{4}}}{\mathcal B} &  e^{-i \theta_{\rm dec}}e^{i  \frac{\pi}{4}} & \frac{e^{i   \frac{\pi}{4}}}{2}\\[4pt]
       \frac{e^{i   \frac{\pi}{4}}}{\mathcal B} e^{- \frac{ \pi \Omega}{\kappa_{\rm eff}}}& e^{-i \theta_{\rm dec}}e^{i  \frac{3\pi}{4}}  e^{\frac{ \pi \Omega}{\kappa_{\rm eff}}} &\frac{e^{-i  \frac{\pi}{4}}}{2}e^{- \frac{ \pi \Omega}{\kappa_{\rm eff}}}
    \end{pmatrix}\,.
\end{align}
Note that the choice made between \cref{eq:Ctilde_VS_C} and \cref{eq:growing} has truly allowed us to write down a third linearly independent combination. This can be easily checked at this level just by noticing that the first and the third columns of $\mathcal U$ are not proportional one another, due to a different phase in the third row.

Computing the determinant of the matrix gives us just a pure phase, as it should be
\begin{align} 
    \det\mathcal U= e^{-i \theta_{\rm dec}}\,.
\end{align}

In order to determine the Bogoliubov coefficients, we proceed as in \cite{Coutant:2011in}, i.e.~we compute a scattering process, where the amplitude of the incoming modes set the initial conditions. 

We choose to start with an \textit{in}-state given by $\varphi_\Omega^{\rm red}$, setting to 0 the amplitude of the negative energy mode $\varphi_\Omega^{\rm or}$. As for the \textit{out}-state, since the incoming mode is bounded, we set to 0 the amplitude associated to the growing one. The other three coefficients can be computed through
\begin{align} 
    \begin{pmatrix}
        \beta_\Omega  \\[4pt]
        1 \\[4pt]
        0
    \end{pmatrix} =\mathcal U \begin{pmatrix}
        \alpha_\Omega  \\[4pt]
        d_\Omega \\[4pt]
        0
    \end{pmatrix}\,.
\end{align}
The coefficients $\beta_\Omega$ and $\alpha_\Omega$ are the Bogoliubov coefficients of the two soft modes. They obey the following relation
\begin{align} 
    \beta_\Omega=e^{i \theta_{\rm dec}} \alpha_\Omega \frac{\mathcal{A}}{\mathcal B} e^{-2 \pi \Omega/\kappa_{\rm eff}}= e^{i \theta_{\rm dec}} \alpha_\Omega e^{-\pi \Omega/\kappa_{\rm eff}} \,,
\end{align}
so that
\begin{align} 
   \biggl| \frac{\beta_\Omega}{\alpha_\Omega} \biggr|^2=e^{-2\pi \Omega/\kappa_{\rm eff}} \,.
\end{align}

Since the decaying mode vanishes at infinity, to conserve the total probability we have to fulfill \cref{eq:completeness}, which brings to an occupation number
\begin{align} \label{eq:Bose_Einstein}
n_\Omega=|\beta_\Omega|^2=\frac{1}{e^{2 \pi \Omega /\kappa_{\rm eff}}-1} \,.
\end{align}
\cref{eq:Bose_Einstein} represents a Bose--Einstein distribution for the emitted phonons, at the same temperature found in \cref{eq:T_efh}. This correspondence emphasize the matching of the tunneling approach with the Bogoliubov coefficient.
Note however, that in passing from this number density to the effective spectrum some grey-body factor will arise even in (1+1) dimensions due to the coupling of modes induced by the modified dispersion relation \cite{Macher:2009tw}. Indeed, these coupling are also those which would induce a partial failing of the completeness relation at high energy (see discussion before \cref{eq:completeness}).

Finally, we recall that \cref{eq:Bose_Einstein} represents the leading order expression for a perturbative expansion of $\beta_\Omega$ in the parameter given by \cref{eq:perturbative_param}. Estimating a perturbative correction $\delta \beta_\Omega$ in terms of $(\bar k/\Lambda)$ can be done by invoking the \qu{broadened horizon paradigm} as in \cite{Coutant:2014cwa}. Indeed the leftmost-hand-side of \cref{eq:both_bounds_2} allows for interpreting \cref{eq:perturbative_param} as a small, local modification of the background flow $\delta v(x) \sim ( k(x)/ \Lambda)^\gamma \ll v(x)$ near to the EFH. Under this modification, one can show, through the distorted wave Born approximation \cite{Newton:1982qc}, that $\beta_\Omega$ would get a correction of the order
\begin{equation} \label{eq:delta_beta}
    \delta \beta_\Omega \sim \int \left(\varphi_\Omega^{\rm H}(x)\right)^* \delta v(x) \varphi_\Omega^{\rm red}(x){\rm d}x \,,
\end{equation}
which is nothing but the linear correction in $\delta v$ to $\beta_\Omega$, coming from modifying the usual Klein-Gordon inner product given by perturbing the background geometry. In the low-energy regime (\cref{eq:perturbative_param}) $|\delta \beta_\Omega| \ll |\beta_\Omega|$, due to the $\Lambda$-suppression of $\delta \beta_\Omega$ and, as shown in \cite{Coutant:2014cwa}, \cref{eq:delta_beta} essentially vanishes in cases where the scale of variation of $\delta v$ is small compared to $1/\kappa_\textsc{efh}$. Namely, the particle production is insensitive to the details of the geometry which happens below the horizon's characteristic length scale.

\section{Estimate of the corrections}

We have seen that the propagation of the modes acquires an explicit $\Omega$-dependence, because of the modified dispersion relation. This dependence is then inherited by the peeling surface gravity $\kappa_\textsc{efh}$ as well as the temperature~\eqref{eq:T_efh}. In this sense, the background geometry acts as a dispersive medium for each frequency individually.

The function $\kappa_\textsc{efh}(\Omega)$ can be studied numerically, as in~\cite{DelPorro:2024tuw}, but it also admits an analytic expansion in the low-energy regime. Working in a perturbative regime, the first correction in $\Omega/\Lambda$ can be computed just by considering a quartic superluminal dispersion relation
\begin{equation}\label{eq:SLquarticDR}
    \omega^2=k^2+\frac{k^4}{\Lambda^2} \,.
\end{equation}
For $\Omega\ll \Lambda$, one finds (see Appendix~\ref{sec:derivation})
\begin{equation}\label{eq:kappa_eff_low_energy}
    \kappa_\textsc{efh}(\Omega)
    =
    \kappa_\textsc{kh}
    \left[
        1+\frac{3}{8}
        \left(
            1-\frac{\kappa'_\textsc{kh}}{\kappa_\textsc{kh}^2}
        \right)
        \frac{\Omega^2}{\Lambda^2}
    \right]
    +\mathcal O\!\left(\frac{\Omega^4}{\Lambda^4}\right) \,,
\end{equation}
where $\kappa_{\rm KH}=v'(0)$ and $\kappa'_\textsc{kh}=v''(0)$ are the Killing-horizon surface gravity and its derivative evaluated at the horizon respectively\footnote{It is noteworthy that the numerical factor of $3/8$ appeared also in khronometric gravity (cf. \cite{DelPorro:2023lbv}) where the leading order in the dispersion relation resembles \eqref{eq:SLquarticDR}.}. Hence, the first dispersive correction is quadratic in $\Omega/\Lambda$, as expected from the dispersion relation by power-counting arguments.

The coefficient $\Omega_{\rm eff}(\Omega)$ controlling the logarithmic singularity in \cref{eq:Omega_eff} can be interpreted as an effective frequency, which admits the low-energy expansion ($\Omega\ll\Lambda$)
\begin{equation}\label{eq:Omega_eff_low_energy}
    \Omega_{\rm eff}(\Omega)
    =
    \Omega
    \left[
        1+\frac{1}{8}\frac{\Omega^2}{\Lambda^2}
    +\mathcal O\!\left(\frac{\Omega^4}{\Lambda^4}\right)\right] \,.
\end{equation}
The low energy expansion of the effective surface gravity $\kappa_{\rm eff}=\kappa_\textsc{efh}/\Omega_{\rm eff}$ combines~\eqref{eq:kappa_eff_low_energy} and~\eqref{eq:Omega_eff_low_energy}.
Therefore, with \eqref{eq:effective_rate} and the relation between the temperature and the effective surface gravity $ T_{\rm eff}(\Omega):=\kappa_{\rm eff}(\Omega)/2\pi$, we find that at low energies the effective temperature contains the two lowest order corrections
\begin{equation}\label{eq:T_eff_low_energy}
    T_{\rm eff}(\Omega)= \frac{\kappa_\textsc{kh}}{2\pi}\left[ 1+\left(\frac{1}{4}-\frac{3}{8}\frac{\kappa'_\textsc{kh}}{\kappa_\textsc{kh}^2}  \right) \frac{\Omega^2}{\Lambda^2} \right]+\mathcal O\!\left(\frac{\Omega^4}{\Lambda^4}\right) \,.
\end{equation}
For a strictly linear near-horizon profile (as in the example considered in \cite{DelPorro:2024tuw}), namely $\kappa'_\textsc{kh}\equiv0$, these expressions simplify to a correction that is only determined by the modified dispersion. This correction is always positive for a superluminal horizon. The additional, non-adiabatic flow correction $\kappa_\textsc{kh}'/\kappa_\textsc{kh}^2$, instead reduces the temperature for a (future) horizon with $\kappa_\textsc{kh}'>0$. 
Therefore, in the superluminal case, the leading dispersive correction raises the effective temperature, although only by a small amount in the regime $\Omega\ll\Lambda$ relevant for the validity of the effective description.

\section{Discussion}
\label{sec:discussion}

{In this work, we set out to test and place on firmer analytical grounds the approximant-based tunneling picture introduced in~\cite{DelPorro:2024tuw}.%
Our main result is that, for a broad class of even, convex, and polynomially bounded, superluminal dispersion relations, the tunneling approach and the Bogoliubov/S-matrix analysis give the same leading particle-production law in the regime of validity of the approximations employed. In particular, both methods identify the relevant outgoing modes as peeling from a dispersive effective horizon rather than from the geometric Killing horizon, and both associate the emission rate to the corresponding peeling coefficient. In this sense, the present analysis provides an explicit analytical bridge between the quasi-local tunneling picture and the more global Bogoliubov-coefficient formalism developed in~\cite{Coutant:2011in}.

The main conceptual payoff of this result is twofold. First, it confirms that the approximant is not just a heuristic shortcut, but a controlled description of the relevant near-horizon outgoing branch whenever the low-energy and adiabatic assumptions hold. Second, it sharpens where and why that construction might fail. The approximant works because the outgoing soft branch can be reconstructed from the regular ingoing mode, whose momentum remains in the low-energy regime and therefore stays under perturbative control. In this regime, the non-analytic structure of the action in momentum space reproduces exactly the logarithmic behaviour that underlies the tunneling result. Conversely, once the conditions selecting the approximant cease to hold, the outgoing branch can no longer be faithfully encoded in terms of the regular ingoing one, and the mismatch should be interpreted not as a breakdown of Hawking radiation itself, but as the breakdown of the simplified effective-horizon description.

This also gives a preciser interpretation of robustness. What is robust is not exact thermality in the strict relativistic sense, but the existence of a universal near-horizon mode-conversion mechanism fully determined by the regularity of the vacuum state and the peeling rate of the relevant outgoing mode. 

More specifically, in dispersive theories, the horizon experienced by the emitted quanta is generically frequency dependent; so is the corresponding peeling surface gravity. As a consequence, the spectrum is not exactly thermal: dispersive corrections induce controlled departures from a pure Planckian law. Nevertheless, our analysis shows that such deviations are themselves governed by a small and calculable set of quantities, and that the leading result is fixed already at the level of the effective horizon.

A particularly suggestive aspect of the derivation is that the whole construction is ultimately anchored in the behaviour of the regular ingoing mode. Once this mode is known, one is able to reconstruct the relevant outgoing branches through the approximant and, from them, infer the Hawking pair. In this sense, the regularity requirement on the vacuum across horizon crossing implies more than the mere selection of an admissible state: under the conditions spelled out in this work, it effectively fixes the mode-conversion pattern responsible for Hawking emission. Put differently, one does not need detailed control over the full high-energy outgoing trajectories to infer the low-energy radiation; the regular infalling sector already contains the essential information. We believe that this observation captures an important part of the physical reason why Hawking radiation remains so robust against ultraviolet modifications.

{Let us also note that the relation between the regular ingoing mode and the Hawking pair becomes exact in the relativistic limit, where the propagation of massless modes is universal. Since this map does not involve trans-Planckian frequencies, one might wonder if it entails the ingoing-outgoing mode-conversion mechanism sometimes conjectured as the true origin of the Hawking quanta \cite{Jacobson:1996zs}. Moreover, as this is a purely kinematical statement --- relating different components of the particles’ four-momenta --- it suggests a connection of this point of view with derivations of Hawking radiation based on Feynman diagrams~\cite{Parentani:1999qv}, in which such kinematical relations appear as conservation laws.} 

{Our treatment also helps disentangle the role of the various approximations. The first is the low-energy condition on the ingoing branch, which ensures that dispersive corrections remain perturbative even though the outgoing partner becomes arbitrarily blueshifted near the effective horizon. The second is the adiabaticity condition, which allows one to identify the coefficient of the logarithmic singularity with the peeling surface gravity and thereby matches the tunneling computation. The third is ignoring the backscattering and the coupling to spectator modes, which is justified in the same low-energy regime and isolates the universal near-horizon mixing from propagation-dependent greybody effects. Taken together, these assumptions define the domain in which the approximant is reliable and in which the Hawking process is analytically under control.

It is worth emphasizing that going beyond this regime should not be viewed as a pure technical complication, but as physically informative. If the approximant-fixing condition is violated, or if the adiabaticity condition fails, the simple effective-horizon picture must be amended. In that case the logarithmic structure in the action may still be present, but its interpretation in terms of a single mode-dependent temperature becomes less direct, and the transfer matrix may mix sectors in a less universal manner. This suggests a natural hierarchy: the existence of Hawking-like emission is the most robust statement; its description in terms of an effective horizon with a single peeling coefficient is the next, more restrictive layer; exact thermality is the special relativistic limit.

Although our analysis focused on superluminal dispersion, the lessons are more general. For subluminal dispersion relations the mode structure is qualitatively different, since the relevant turning points and asymptotic branches are rearranged, and the effective horizon construction may need to be reformulated accordingly. Likewise, dissipative or non-Hamiltonian ultraviolet modifications would require a further conceptual extension, because the very notions of conserved norm, transfer matrix, and WKB phase acquire a different status. Still, the present result suggests that the key question in those cases will again be whether one can identify a regular low-energy infalling sector and a controlled mechanism relating it to the outgoing modes. We hope to return to these issues elsewhere.

To summarize, the present work strengthens the case for the robustness of Hawking radiation in dispersive media by showing that the tunneling and Bogoliubov approaches are not competing descriptions, but two complementary ways of accessing the same near-horizon physics. Quantitatively, we recover the dispersive corrections anticipated in the tunneling approach and place them within the S-matrix framework. Qualitatively, we learn that the essential input is the propagation of the regular ingoing mode and the vacuum regularity it encodes. Once these are under control, Hawking emission follows under broad conditions, even though exact thermality need not.}

\begin{acknowledgments}
The authors wish to thank Stefano Finazzi and Ralf Schützhold for insightful discussions. The research of FDP is supported by a research grant (VIL60819) from VILLUM FONDEN. The Center of Gravity is a Center of Excellence funded by the Danish National Research Foundation under grant No. 184. The work of MS has been supported by the Basque Government Grant
\mbox{IT1628-22} as well as by the Grant PID2021-123226NB-I00 (funded by
MCIN/AEI/10.13039/501100011033 and by ``ERDF A way of making Europe'')
\end{acknowledgments}

\appendix

\section{Derivation $\kappa_{\textnormal{\textsc{efh}}}$ in terms of $\kappa_{\textnormal{\textsc{kh}}}$}
\label{sec:derivation}
To derive the low-energy expansion of the effective surface gravity $\kappa_\textsc{efh}$, let us start from its definition
\begin{equation}
    \kappa_\textsc{efh}:=\left.\frac{\dd}{\dd x}\Bigl(v(x)-c_g^-(x,\Omega)\Bigr)\right|_{x=x_0},
\end{equation}
where $x_0$ marks the location of the effective horizon. It is determined by the condition
\begin{equation}
    v(x_0)=c_g^-(x_0,\Omega)=F'(\bar k(x_0,\Omega)),
\end{equation}
where $\bar k(x_0,\Omega):=k_-(x_0,\Omega)$ and the prime denotes the partial derivative with respect to the argument. Since by definition $c_g^-(x,\Omega)=F'(k_-(x,\Omega))$, the surface gravity becomes
\begin{equation}\label{eq:kappa_eff_step1}
    \kappa_\textsc{efh}=v'(x_0)-F''(\bar k)\,k_-'(x_0,\Omega).
\end{equation}
We now compute $k_-'(x_0,\Omega)$ by differentiating the ingoing dispersion relation
    $\Omega-v(x)k_-(x,\Omega)=F(k_-(x,\Omega))$
with respect to $x$ at fixed $\Omega$. Evaluating at $x=x_0$ and using $v(x_0)=F'(\bar k)=\bar v$, we find for $k'_-(x_0,\Omega)$
\begin{equation}\label{eq:kprime_x0}
    k_-'(x_0,\Omega)=-\frac{v'(x_0)\bar k}{2\bar v}.
\end{equation}
Substituting~\eqref{eq:kprime_x0} into~\eqref{eq:kappa_eff_step1}, we obtain
\begin{equation}\label{eq:kappa_eff_master}
    \kappa_\textsc{efh}=v'(x_0)\left(1+\frac{F''(\bar k)\bar k}{2\bar v}\right).
\end{equation}

Let us now specialise to the quartic superluminal dispersion relation. For the exterior ingoing solution, the momentum $\bar k<0$.

A straightforward computation then gives with $\bar v=v(x_0)=F'(\bar k)$ and the replacement $y=\frac{\bar k^2}{\Lambda^2}$
\begin{equation}\label{eq:Fexpand}
    \frac{F''(\bar k)\bar k}{2\bar v}=\frac{y(3+2y)}{2(1+y)(1+2y)}.
\end{equation}

To obtain the low-energy expansion, we must now express both $v'(x_0)$ and $y$ perturbatively. Expanding the flow profile around the geometric Killing horizon at $x=0$, we write
\begin{equation}\label{eq:v_expand_kappa}
    v(x)=-1+\kappa_\textsc{kh}x+\frac12 \kappa'_\textsc{kh}x^2+\mathcal O(x^3),
\end{equation}
where $\kappa_\textsc{kh}:=v'(0)$ and $\kappa'_\textsc{kh}:=v''(0)$. %Recall that the effective horizon is defined by $v(x_0)=F'(\bar k)$, 
Thus by using
\begin{equation}
    F'(y)= -\frac{1+2y}{\sqrt{1+y}}=-1-\frac32 y+\mathcal O(y^2),
\end{equation}
together with~\eqref{eq:v_expand_kappa}, we obtain
\begin{equation}
-1+\kappa_\textsc{kh}x_0+\mathcal O(x_0^2)=-1-\frac32 y+\mathcal O(y^2).
\end{equation}
We fixed the sign of $F(\bar k)$ such that $v(0)=-1$. Since the displacement of the effective horizon $x_0=\mathcal O(y)$, this implies $   x_0 =-\frac{3}{2}\frac{y}{\kappa_\textsc{kh}} +\mathcal O(y^2)$.
Hence, for small energies, we can rewrite the effective horizon's surface gravity as the Killing horizon's surface gravity plus corrections
\begin{align}
    v'(x_0)=\kappa_\textsc{kh}+\kappa_\textsc{kh}'x_0+\mathcal O(x_0^2)= \kappa_\textsc{kh}\left[   1-\frac{3}{2}\frac{\kappa_\textsc{kh}'}{\kappa_\textsc{kh}^2}y \right]+\mathcal O(y^2).\nonumber
\end{align}
On the other hand, when expanding $\kappa_\textsc{efh}$ at small values of $y$ using \eqref{eq:kappa_eff_master} and \eqref{eq:Fexpand} and combining this with the previous expansion yields
\begin{align}
    \kappa_\textsc{efh} =\kappa_\textsc{kh} \left(1-\frac{3}{2}\frac{\kappa_\textsc{kh}'}{\kappa_{\rm KH}^2}y\right)\left(1+\frac32 y \right)+\mathcal O(y^2)=\kappa_\textsc{kh}\left[1+\frac32\left(  1-\frac{\kappa_\textsc{kh}'}{\kappa_{\rm KH}^2} \right)y \right]+\mathcal O(y^2).
\end{align}
Finally, using the condition for the effective horizon
\begin{equation}
    \Omega=F(\bar k)+\bar k F'(\bar k)=-\bar k\,\frac{2+3y}{\sqrt{1+y}}= 2|\bar k|+\mathcal O(\bar k \,y),
\end{equation}
where in the last step, we employed $\bar k<0$, one gets $ y=\frac{\Omega^2}{4\Lambda^2}+\mathcal O\!\left(\frac{\Omega^4}{\Lambda^4}\right)$.
Substituting this into the previous expression yields
\begin{equation}
    \kappa_\textsc{efh}(\Omega) = \kappa_\textsc{kh}  \left[1+\frac{3}{8} \left( 1-\frac{\kappa_\textsc{kh}'}{\kappa_{\rm KH}^2}\right)\frac{\Omega^2}{\Lambda^2} \right]+\mathcal O\!\left(\frac{\Omega^4}{\Lambda^4}\right),
\end{equation}
which yields the desired result.

\bibliography{apssamp}% Produces the bibliography via BibTeX.

\end{document}